# Binospec: A Wide-field Imaging Spectrograph for the MMT

**Daniel Fabricant[1], Robert Fata[1], Harland Epps[2], Thomas Gauron[1], Mark Mueller[1], Joseph Zajac[1], Stephen Amato[1], Jack Barberis[1], Henry Bergner[1], Patricia Brennan[1], Warren Brown[1], Igor Chilingarian[1], John Geary[1], Vladimir Kradinov[1], Brian McLeod[1], Matthew Smith[1], Deborah Woods[3]**

[1] Harvard-Smithsonian Center for Astrophysics, Cambridge, MA, USA
[2] Lick Observatory, Santa Cruz CA, USA
[3] MIT Lincoln Laboratories, Lincoln MA, USA

E-mail: dfabricant@cfa.harvard.edu

**Abstract**

Binospec is a high throughput, 370 to 1000 nm, imaging spectrograph that addresses two adjacent 8′ by 15′ fields of view. Binospec was commissioned in late 2017 at the f/5 focus of the 6.5m MMT and is now available to all MMT observers. Aperture masks cut from stainless steel with a laser cutter are used to define the entrance apertures that range from 15′ long slits to hundreds of 2″ slitlets. System throughputs, including the MMT's mirrors and the f/5 wide-field corrector peak at ~30%. Three reflection gratings, duplicated for the two beams, provide resolutions ($\lambda/\Delta\lambda$) between 1300 and >5000 with a 1″ wide slit. Two through-the-mask guiders are used for target acquisition, mask alignment, guiding, and precision offsets. A full-time Shack-Hartmann wave front sensor allows continuous adjustment of primary mirror support forces, telescope collimation and focus. Active flexure control maintains spectrograph alignment and focus under varying gravity and thermal conditions.

## 1. Introduction

The f/5 focus of the 6.5m MMT telescope was designed to support wide-field faint-object spectroscopy (Fabricant et al. 1994, Fata and Fabricant 1993). The full 1° diameter field of view is addressed with optical fibers in Hectospec/Hectochelle (Fabricant et al. 2005, Szentgyorgyi et al. 2011). Our goal for Binospec is to address the largest possible field with a direct imaging spectrograph. Binospec uses the telecentric 1° field mode of the f/5 wide field corrector that provides atmospheric dispersion compensation. The relative merits of optical fiber and direct spectrographs are hotly debated, but direct spectrographs offer maximum flexibility in the configuration of the entrance aperture, higher throughput, and superior sky subtraction, usually at the cost of field of view. We consider Binospec to be a complement to rather than a replacement for the MMT's fiber-fed spectrographs for multi-object spectroscopy. Binospec joins a cohort of capable wide-field direct spectrographs in operation at 8m class telescopes including DEIMOS at Keck (Faber et al. 2003), GMOS at Gemini (Hook et al. 2004), IMACS at Magellan (Dressler et al. 2011), MODS at LBT (Pogge et al. 2006), and FORS at VLT (Appenzeller et al. 1998).

As its name suggests, Binospec has two beams. In many cases dual beam spectrographs are used with dichroics to split the spectrum into red and blue channels. While this approach has advantages, Binospec's two beams are identical in design, and address two adjacent 8′ by 15′ regions on the sky. The narrower 8′ dimension is parallel to the dispersion direction, allowing a total slit length of up to 15′. The gap between the 8′ wide entrance apertures of the two beams is 32 mm, or 3.2′. A pair of entrance aperture masks are mounted on a common frame, and ten mask frames can be mounted at one time.

Six interchangeable filters are deployed below the entrance aperture, before the first collimator lens. The filters for both beams are mounted on a common frame. After this point, all optics and mechanisms for the two beams are duplicated. Each beam has an independent mechanism for exchanging and adjusting the incidence angles of imaging mirrors or diffraction gratings. Binospec accommodates four mirrors or gratings for each beam. The dispersed light from each beam is imaged onto a single 4K by 4K CCD, and focal plane shutters are deployed just before the dewar entrance windows.

Binospec has attracted a diverse group of users employing all of Binospec's capabilities. In the first four months of 2019, 18 programs have been approved for observations ranging from exoplanet atmospheres to z=5-7 galaxies. In this period just over half the programs use multislit masks and Binospec's full field of view, while the remainder are mostly long slit observations. The long slit observations are dominated by observations of transients, although high redshift quasars are also popular targets. There are a smattering of imaging observations, typically of transients as well. The first paper published with Binospec data was the discovery of a z=6.5 lensed quasar (Fan et al. 2019).

Binospec is operated entirely in queue mode by observers who also operate the other MMT f/5 instruments in queue mode (Hectospec, Hectochelle, and MMIRS). The queue mode allows efficient scheduling of targets of opportunity as well as time critical observations, e.g. exoplanet transits. Queue observing also allows more efficient scheduling of conventional observations than is usually possible with a single program using an entire night and distributes the risk of bad weather. Queue observing requires more software infrastructure than conventional observing but the rewards are considerable.

Here we describe Binospec's hardware, including optics, structure, mechanisms. In a companion paper (Kansky et al. 2019), we describe Binospec's control, operation, observation planning and data reduction software.

## 2. Optical Design, Analysis, and Performance

### *2.1 Optical Layout*

Each Binospec beam includes a refractive collimator and a refractive camera, with focal lengths 1097 mm and 404 mm, respectively, producing a scale demagnification of 2.7, and a scale at the detector of 0.24″ per 0.015 mm pixel. Binospec's collimated beam diameter is ~208 mm. Figure 1 shows the Binospec optical layout with all of the fold mirrors and grating removed.

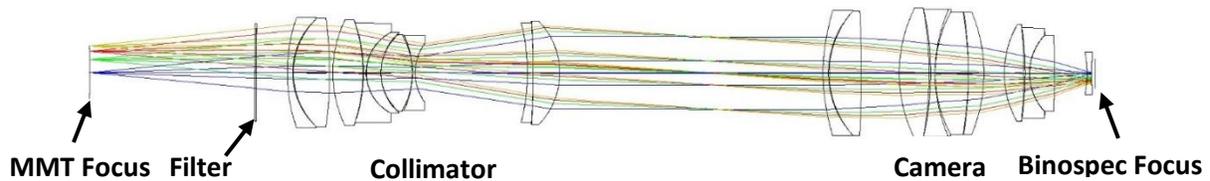

**Figure 1. The Binospec optics with all fold mirrors removed. The total path length from the entrance aperture on the left and the detector on the right is 2.9 m. The collimator has three lens groups and the camera has four including the field flattener/dewar window lens.**

Figure 2 shows Binospec's three-dimensional optical layout, from the entrance aperture at the top to the detector at the lower right. To maintain a compact package, Binospec uses three fold mirrors in the collimator as well as reflection gratings (or an imaging mirror). The first two mirrors act as a periscope to separate the two beams to accommodate the collimator optics side by side. The third fold mirror, following the quintet in the collimator, folds the beam so that the subsequent optics can all sit in a plane on an optical bench.

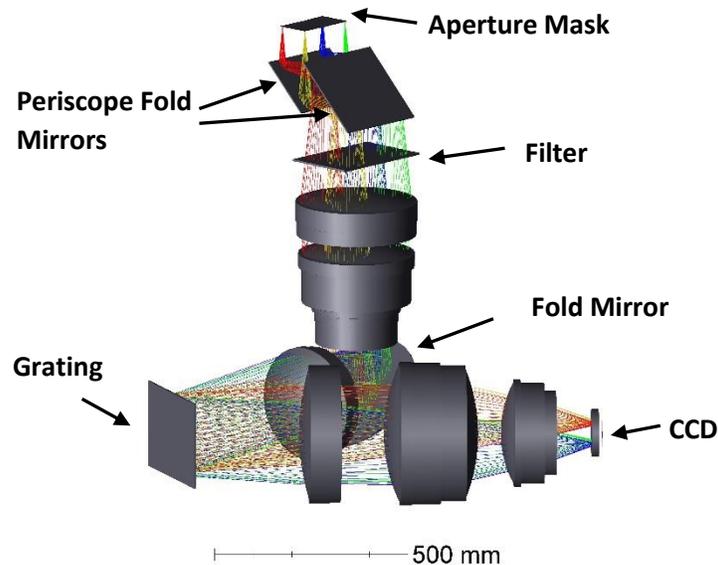

**Figure 2. A three-dimensional solid model of one of Binospec's two beams.**

### *2.2 Optical Design*

The optical design goals were to maintain high throughput and excellent image quality over the 370 to 1000 nm spectral range. The design therefore used Ohara i-line glasses and crystalline materials ($CaF_2$ and $NaCl$). The authors greatly expanded their skill set learning to work with the very hygroscopic $NaCl$, but cannot recommend this experience to others absent exceptional motivation. These materials, as well as the LL5610 coupling fluid, are essentially perfectly transparent in the Binospec bandpass, and the only significant losses arise from the antireflection coating at glass-air surfaces.

All of the multiplets are coupled internally with Cargille Laboratories LL5610 with a refractive index $n_D = 1.5$. Although containing the fluid proved a significant challenge, the throughput advantage of avoiding throughput losses at glass-air surfaces is strong motivation. The worst index match at the D line (589.3 nm) is to the PBM2Y glass, with $n_D=1.620$, resulting in a reflection loss of 0.15% at this surface. A broadband antireflection coating with this level of performance is impossible to achieve with present technology. A more typical index mismatch of 0.06 results in a reflection loss four times smaller.

Binospec uses reflection gratings partially because they provide a convenient fold in the optical path. Volume phase holographic (VPH) gratings are now commonly used in optical spectroscopy because they offer a higher peak throughput than reflection gratings. However, VPH gratings more than give back this advantage at the edges of the field in a wide-field spectrograph because the Bragg condition is no longer satisfied.

With three fold mirrors in the collimator, we put considerable energy into identifying coatings with higher reflectivity than standard metallic coatings. Various forms of silver coatings, protected and enhanced in the UV with overcoats, are available but in most cases offer unknown lifetime. We elected to use a dielectric stack developed by Evaporated Coatings, Inc. This remarkable coating offers >99% reflectivity (averaged over polarizations) from 400 to 1050 nm, dropping to 97% at 380 nm.

**Table 1. Binospec optical prescription (one of two beams). A positive radius has a convex side facing the telescope focal surface. Surfaces 7-10 form a doublet, collimator group 1. Surfaces 11-20 form a quintet, collimator group 2. Surfaces 22-25 form a doublet, collimator group 3. Surfaces 27-30 form a doublet, camera group 1. Surfaces 31-38 form a quartet, camera group 2. Surfaces 39-44 form a triplet, camera group 3. All multiplets are coupled with Cargille Laboratories LL5610 fluid. The lens groups are set off in alternate shades of grey. All optical glass lenses are Ohara i-line glass, the $CaF_2$ was supplied by Canon, and the NaCl was supplied by Hilger Crystals. The prescription is exact for the specific glass melts at 24 °C.**

| Surface | Radius | Thickness | Material | Diameter | Conic | Notes |
|---|---|---|---|---|---|---|
| 1 | -3404.0 | -0.445 | | | -665 | Telescope focal surface |
| 2 | - | 43.129 | | | | X offset 56.235 mm, chief ray enters at 0.72° tilt about Y |
| 3 | Infinity | 151.528 | Mirror | 165 x 228 | 0 | 91.44° fold relative to chief ray, X offset 7.2 mm |
| 4 | Infinity | 181.042 | Mirror | 226 x 250 | 0 | 90.72° fold (chief ray exits parallel to telescope axis), X offset 8.2 mm |
| 5 | Infinity | 5.000 | N-BK7 | 185 x 280 | 0 | Filter |
| 6 | Infinity | 88.843 | | 185 x 280 | 0 | |
| 7 | 493.081 | 15.365 | BAL15Y | 316 | 0 | [1]Asphere, $9.543711 \times 10^{-10}$, $6.177683 \times 10^{-15}$, $3.805871 \times 10^{-19}$ |
| 8 | 209.254 | 2.375 | LL5610 | 316 | 0 | |
| 9 | 213.794 | 104.140 | S-FSL5Y | 311 | 0 | |
| 10 | -1113.050 | 9.340 | | 311 | 0 | |
| 11 | 348.371 | 86.520 | PBM2Y | 302 | 0 | |
| 12 | -454.440 | 1.671 | LL5610 | 302 | 0 | |
| 13 | -429.230 | 7.600 | PBL6Y | 284 | 0 | |
| 14 | 129.379 | 0.677 | LL5610 | 284 | 0 | |
| 15 | 129.372 | 41.792 | BAL35Y | 231 | 0 | |
| 16 | 153.080 | 0.249 | LL5610 | 231 | 0 | |
| 17 | 145.199 | 90.262 | $CaF_2$ | 219 | 0 | |
| 18 | -214.602 | 0.259 | LL5610 | 219 | 0 | |
| 19 | -250.067 | 7.887 | PBL6Y | 214 | 0 | |
| 20 | 142.940 | 163.830 | | 214 | 0 | |
| 21 | Infinity | 158.869 | Mirror | 303 | 0 | 90° fold |
| 22 | -646.289 | 16.787 | BSM51Y | 286 | 0 | [1]Asphere, $-3.705514 \times 10^{-9}$, $-3.471690 \times 10^{-14}$, $-6.176305 \times 10^{-19}$ |
| 23 | -2223.477 | 0.259 | LL5610 | 286 | 0 | |
| 24 | -2493.898 | 76.297 | $CaF_2$ | 296 | 0 | |
| 25 | -211.860 | 431.800 | | 296 | 0 | |
| 26 | Infinity | 326.390 | Mirror | varies | 0 | 45° fold – mirror or diffraction grating |
| 27 | 546.538 | 15.443 | BAL35Y | 358 | 0 | [1]Asphere, $-2.147458 \times 10^{-9}$, $-1.043657 \times 10^{-14}$, $-1.369504 \times 10^{-19}$ |
| 28 | 260.022 | 2.079 | LL5610 | 358 | 0 | |
| 29 | 265.437 | 92.650 | $CaF_2$ | 357 | 0 | |
| 30 | -3206.180 | 109.021 | | 357 | 0 | |
| 31 | 416.320 | 86.398 | $CaF_2$ | 372 | 0 | |
| 32 | -1000.978 | 4.300 | LL5610 | 372 | 0 | |
| 33 | -805.380 | 12.725 | BAL35Y | 365 | 0 | |
| 34 | 453.135 | 0.259 | LL5610 | 365 | 0 | |
| 35 | 368.688 | 95.236 | $CaF_2$ | 351 | 0 | |
| 36 | -978.641 | 0.257 | LL5610 | 351 | 0 | |
| 37 | -1156.802 | 44.519 | PBM2Y | 351 | 0 | |
| 38 | -455.303 | 58.816 | | 351 | 0 | |

| 39 | 280.345 | 56.440 | FPL51Y | 277 | 0 |
| 40 | -1337.540 | 3.469 | LL5610 | 277 | 0 |
| 41 | -911.850 | 12.390 | NaCl | 260 | 0 |
| 42 | 159.190 | 0.608 | LL5610 | 260 | 0 |
| 43 | 159.059 | 70.965 | CaF$_2$ | 218 | 0 |
| 44 | -4250.275 | 94.703 |  | 218 | 0 |
| 45 | -273.685 | 11.617 | BSM51Y | 122 | 0 |
| 46 | 337.907 | 11.545 |  | 122 | 0 |
| 47 | Infinity |  |  | 83.13 | 0 |

[1] Aspheric coefficients for 4th, 6th, 8th order terms

## 2.3 Optics Athermalization

The LL5610 has an additional role in Binospec's optical design: it is used to athermalize the optics to operate over a wide temperature range of -10 C to 25 C.  When we analyzed the thermal behavior of the early Binospec optical designs, we discovered that an image at the edge of the field of view would be displaced by one pixel per degree C due to the focal length changes of the optics, predominantly the collimator.  Harland Epps suggested that we allow the coupling fluid interfaces to have a small amount of power.  Because the LL5610 has a large negative dn/dT, these weak fluid lenses can be used to athermalize the optics (Epps and Fabricant 2002). Making this athermalization work in practice requires careful analysis and a detailed knowledge of the thermal properties of all the optical materials and the optics mounts.   We were unable to test the results of this analysis until we commissioned Binospec in 2017, but were gratified to find that the axial position of the focus changed by <3 µm per degree C at the detector, and the lateral scale changes are insignificant.

## 2.4 Optical Testing and Performance

### 2.4.1 Subassembly Optical Tests

The assembled collimators (Figure 3) were optically tested before integration into Binospec.  The collimators were tested in double pass, illuminated by an optical fiber behind a 10 µm pinhole placed on a curved focal plate, and the return image detected by a small USB camera placed on the same focal plate.   A mirror with a pupil stop to define the correct f/5 focal ratio was placed at the collimated light output of the collimator to return the light for double pass operation. Ray traces of a perfect double pass collimator tested in the g, r, i, and z bands predicted RMS image radii in the 17 to 35 µm range in these bands over the entire field.  The first test images from the beam 1 collimator revealed asymmetries not present in the ray trace analysis, but further analysis indicated a lateral misalignment of the third lens group with respect to the first two groups of  ~75 µm and ~50 µm in the two axes.  Small alignment errors in the position and angle of the fold mirror were the likely causes of this error.  Recentering the third lens group eliminated the image asymmetries and yielded RMS images in 20 to 44 µm (radius) range in the four filters over the entire field.  Although these images are slightly worse than predicted, single pass and demagnified by the camera optics, the differences were judged to be insignificant.  Errors in the somewhat complex test setup may also have contributed to these differences.  The second collimator proved to have similar lens group three alignment errors that were also corrected, and very similar optical performance after realignment.  The cameras have no fold mirrors so we did not expect to see comparable alignment issues.  We briefly tested the beam 1 collimator with the beam 2 camera, feeding the system with the same optical fiber in single pass. The small USB camera was mounted on a three axis stage to probe the image quality at the 62 mm by 62 mm final focus. We saw RMS images of 5 to 7.5 µm radius across the field at best focus, but the difficulties of properly aligning the stage system to the camera led us to defer further testing until the optics were integrated into Binospec and we could use the full size science detectors.

*2.4.2 System Optical Tests*

Once the optics were integrated into Binospec we could use calibration slit masks with a regular pattern of 12 µm pinholes for imaging tests or a pattern of 85 µm slits for spectroscopic tests. With broad band g, r, i, or z filters we measured RMS image diameters of 1.24 to 1.28 pixels (~19 µm, or ~0.3″ on the sky), averaged across the 8′ by 15′ field of view in both beams. The worst images at the edge of the field of view are ~15% larger. Spectra of the HeAr calibration lamps obtained with 85 µm (0.5″) wide slits for all three gratings produce slit images of 2.2 to 2.3 pixels (~34 µm or 0.55″) FWHM in both beams, averaged over the field of view.

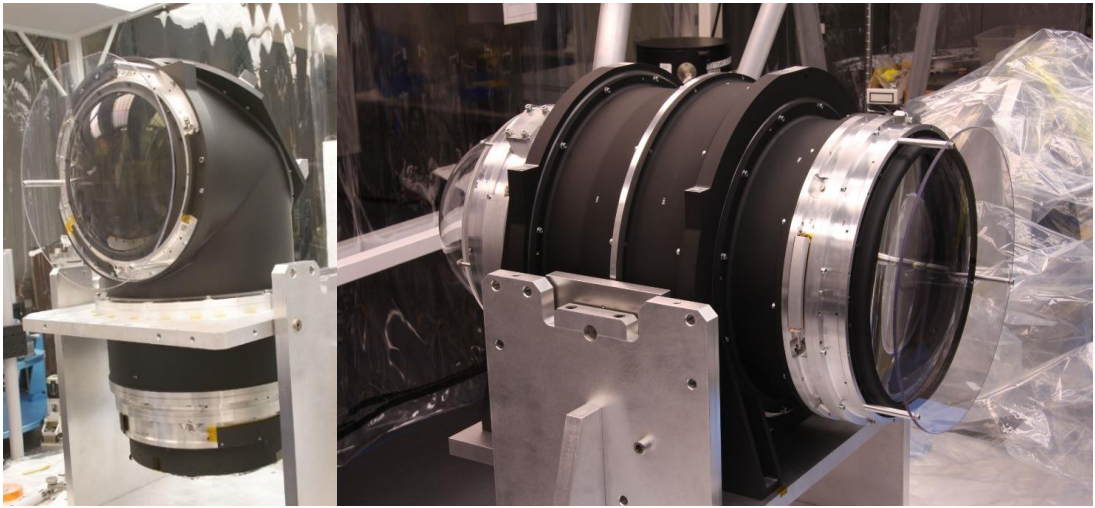

**Figure 3.** (Left) Collimator and (Right) camera lens assemblies for one of Binospec's two beams.

## 3. Making the Optics a Reality

*3.1 Aligning Lenses in their Bezels*

Fata, Kradinov and Fabricant (2006) describe the flexure mounts used for Binospec's camera and collimator lenses (Figure 4). Each lens has an individual bezel, and the bezels are bolted together to form collimator and camera assemblies. Mounting the lenses to their flexures in the lens bezels requires precision alignment in five degrees of freedom: tip, tilt, and position along three axes. Since the optical center of the lens is not exactly the mechanical center, the most precise alignment requires optical detection of the lens axis. This was accomplished with an Opticentric machine.

The Opticentric uses two autocollimators to view both surfaces of a lens simultaneously (Figure 5). The lens and bezel are supported on a precision rotary air bearing. The autocollimators allow us to simultaneously align the front and rear lens vertices to the air bearing rotary axis. The bezel is aligned to the bearing rotary axis with axial and radial dial indicators. The Opticentric is also used to set the height of the lens with respect to the bezel, using a custom fixture that has a small optical flat mounted at its top and that rests on the bezel's axial mounting surface. We achieved axial position and tilt tolerances of the lens relative to the bezel of 21 µm and 40 µrad respectively, within our tolerances of 25 µm and 60 µrad, including the bezel machining tolerances.

Once the lens is aligned to its bezel, it is bonded in place with Hysol 9313 epoxy filled with 40% Siltex silica powder. The epoxy is injected through 1.8 mm diameter holes in the nubs, filling the 0.3 mm gap between the nub and lens.

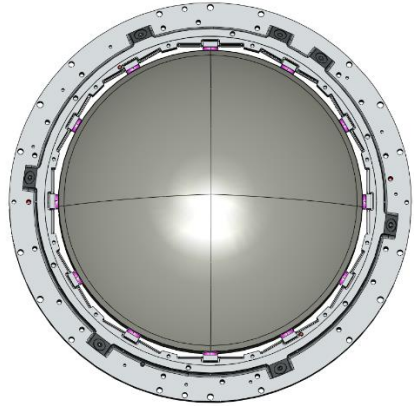

**Figure 4. CAD model of Binospec lens in its bezel, held in place by 12 tangential flexures.  Nubs matched to the coefficient of thermal expansion of the optical glass are shown in purple.**

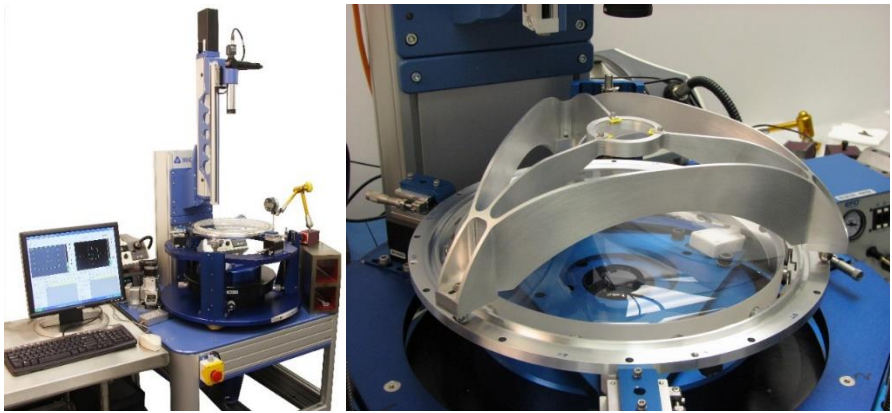

**Figure 5.  (Left) Opticentric lens alignment machine with two autocollimators and a precision air rotary bearing.  (Right) Custom fixture used to measure the axial position of the bezel.**

*3.2 Fold Mirrors*

The first periscope fold mirrors are located directly beneath the slit mask (Figure 6).  Mirrors for both beams are aligned and bonded to a common central aluminum mount using three discrete 15mm diameter x 0.5mm thick RTV60 pads. We used a coordinate measuring machine to align the mirrors to precision reference surfaces machined on the mount.  The second periscope mirrors and the collimator fold mirrors are supported on three bipod flexures bonded to their back surface with Siltex-filled Hysol epoxy. These mirrors were aligned to their mounts on the Opticentric machine in a fashion similar to the lens-bezel alignment described above (see Figure 7). The fundamental frequencies of the 1st, 2nd and 3rd fold mirror mounts are 141, 240 and 250 Hz, respectively and the out-of-plane surface deformations resulting from combining gravity loads and a 40°C temperature change are 0.76, 0.30 and 0.15 µm for the three fold mirrors respectively. These are relatively benign low spatial frequency errors that have a negligible effect on the image quality.

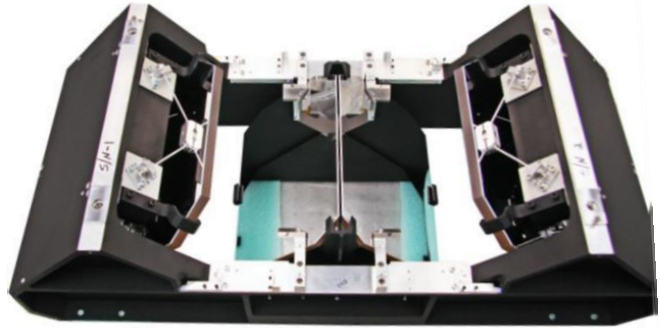

**Figure 6. Periscope fold mirror assembly. This assembly is mounted just behind the slit mask assembly, and bolts to Binospec's focal plane bench. The rear surfaces are coated with the same reflective dielectric stack to balance the stresses on both sides of the mirror to limit thermal deformation of the optical surface.**

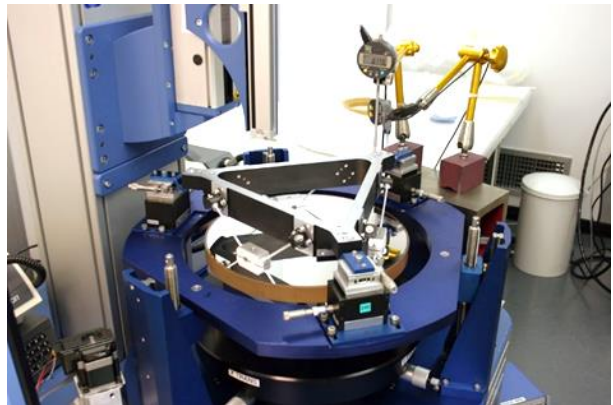

**Figure 7. The collimator fold mirror alignment on the Opticentric.**

### 3.3 Handling the NaCl lenses

The most important issue in handling NaCl optics is maintaining a dry environment. We added a desiccant dehumidifier to the small clean room containing our Opticentric to maintain a relative humidity of <15% for long periods. Inside this room the NaCl lens could be handled freely. We also installed a desiccant storage cabinet capable of maintaining relative humidity <1% indefinitely.

The NaCl lenses are mounted by preloading reference surfaces on the lenses against axial and radial hard points. The diameter of the NaCl lens blanks was minimized to ease generation of lens blanks from the parent crystal. Consequently, the lenses lacked axial mounting features. We bonded NaCl axial mounting ears to the lenses using Hysol 9313 filled with 40% Siltex. Figure 8 shows the mounting ears aligned to the NaCl lens on the Opticentric and the completed lens assembly.

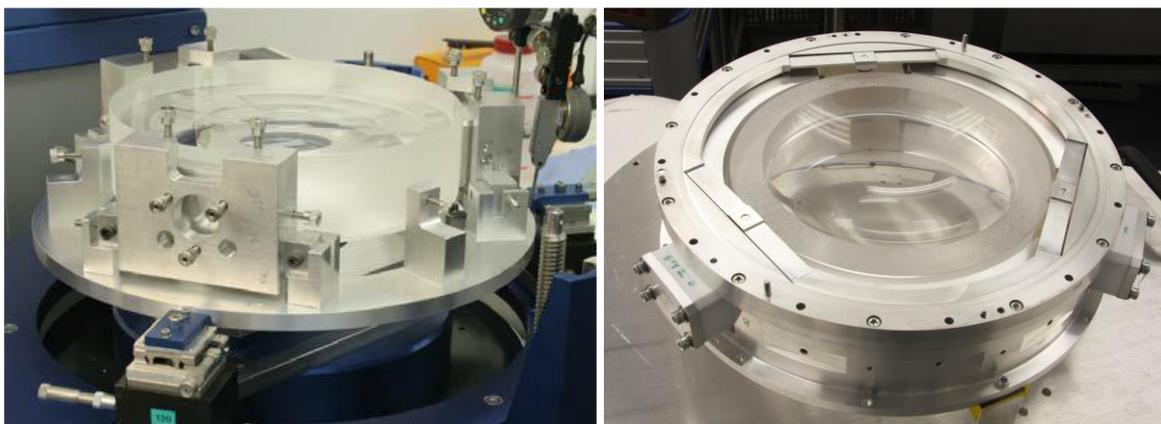

**Figure 8. (Left) Fixture used to bond mounting ears on a cylindrical NaCl lens. (Right) Completed NaCl lens assembly with a centration tolerance of 28 μm and axial placement tolerance of 40 μm.**

*3.4 Working with LL5610 Fluid Couplant*

The seals for the fluid coupled multiplets are challenging because LL5610 is a siloxane compound that reacts with many materials. All materials that potentially contact the LL5610 have been tested to make sure that they do not contaminate the couplant and that the couplant does not degrade their mechanical properties. Material samples are immersed in vials of LL5610 for a least a month before testing begins. Couplant contamination is detected by changes in the LL5610 refractive index or transmission. Cargille Laboratories used a Beckman DK2A spectrophotometer to perform a differential transmittance measurement over the 220 to 2500 nm band for a 1 cm path for a potentially contaminated LL5610 sample relative to a fresh LL5610 sample. A material is judged compatible if the transmission curves match to 1% or better at all wavelengths. The refractive index of the potentially contaminated LL5610 was read on a Bellingham and Stanley 60/LR refractometer at 25 ºC at 589.3 nm where a pure sample has a refractive index of 1.500. A NIST traceable standard was used with a refractive index of 1.51432 at 589.3 nm at 25 ºC. A material is deemed compatible if the LL5610 index did not change more than 0.0001 with respect to witness samples. Compatible materials are listed in Table 2 and incompatible materials are listed in Table 3.

**Table 2. Materials compatible with LL5610**

| |
|---|
| Aluminum |
| Dupont Kapton 500HPP-ST (Corona Treated) |
| Dupont 500JP Formable Polyimide |
| Sodium Chloride |
| Viton |
| Hysol 9313 epoxy filled 40% by weight with Siltex 44 silica powder |
| Teflon FEP (Fluorinated Ethylene Propylene) |
| Solvay Adjedium Films Aurum (thermoplastic polyimide) |
| SABIC Innovative Plastics Ultem (thermoplastic polyetherimide) |

**Table 3.  Materials incompatible with LL5610**

| Material | Issue |
|---|---|
| Momentive (formerly GE) RTV655 two-part clear silicone elastomer | Index change to 1.5133, transmission change |
| Dow Corning Sylgard RTV184 two-part clear silicone elastomer | Index change to 1.5137, transmission change |
| Momentive (formerly GE) RTV560 two-part red silicone elastomer | Decrease in durometer 55 to 30, discoloration observed |
| Buna-Nitrile O-ring | Transmission change |

*3.5 Seals and Expansion Bags*

Working with LL5610 presents the obvious issue of seals, but its volume coefficient of thermal expansion ($8 \times 10^{-4}$ $°C^{-1}$), an order of magnitude larger than aluminum, presents additional challenges.  The decrease in couplant volume as the lens assembly cools would create a vacuum unless some form of volume compensation is provided.  We installed three expansion bags spaced equally around the periphery of the lens multiplets to provide this compensation.

We originally developed seals and expansion bags formed from 0.19 mm thick Teflon because it is easy to form and initially showed no reaction with the Cargille 5610 laser liquid.  The Teflon was etched with sodium ammonia from Porter Process Inc. to provide a bondable surface.  The seals were bonded to the substrate with Hysol 9313 mixed with Siltex 44 silica powder.  The adhesion was initially satisfactory but we observed a drastic decrease in the bond strength after soaking the samples in LL5610 for a month.

We evaluated two alternate candidates to the Teflon, Aurum (thermoplastic polyimide) and Ultem (a thermoplastic polyetherimide), and chose 0.18 mm Aurum because it can be more readily formed into the required seal shape.  We experimented with several Aurum pretreatments including caustic and plasma etches and sand blasting, as well as aluminum and titanium coatings.  The best adhesion is obtained with no coating or surface preparation, with a peel strength of 2.6 N $mm^{-1}$.  The maximum peel load applied to the seals is ~0.09 N $mm^{-1}$.  The bond shear tests resulted in failure of the 0.18 mm Aurum at ~92 MPa before bond failure.  The maximum shear load applied by a 35 °C temperature change and a 3 g gravity load is ~10 MPa.  We also verified that the same peel and sheer strength is achieved after the Aurum has been heat cycled during the seal molding process.

Figure 9 shows the seal shapes for collimator lenses 8 and 9. This shape minimizes the seal load on the lenses.  We use two seal profiles depending on where the seal is bonded to the lens. The length of the seal between the lens and the bezel determines the seal compliance.  The seals are molded into their required shape with a positive and negative mold pair (Figure 10). The Aurum is placed between the mold pairs and the sandwich is heated to 290 °C.  Seals produced in this fashion have uniform thickness and display ~0.5% shrinkage relative to the mold dimensions.

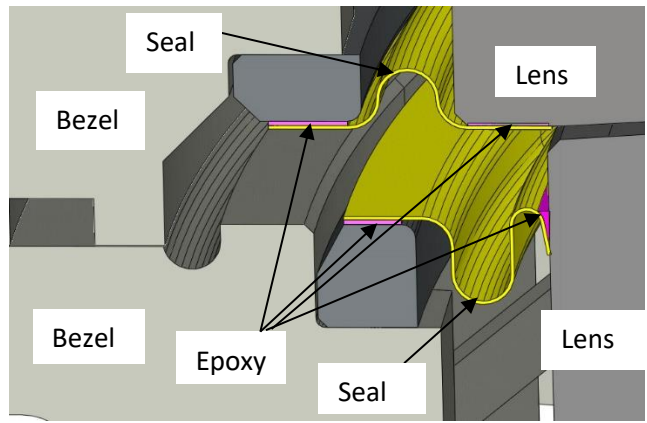

**Figure 9.  Detailed view of the lens seal geometry.**

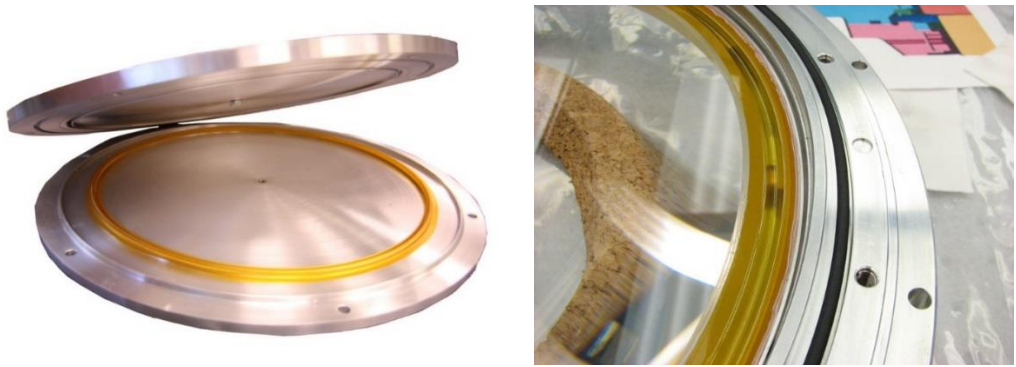

**Figure 10.  A seal mold and a finished seal bonded to a lens.**

Figure 11 shows typical collimator and camera couplant expansion bags. The expansion bags were sized by calculating the difference in fluid volume at -18 °C and 20 °C and adding a safety factor of two.  The collimator bags were heat sealed around the perimeter except for a small hole at the end.  An aluminum tube was bonded into the hole with Siltex filled Hysol 9313 epoxy.

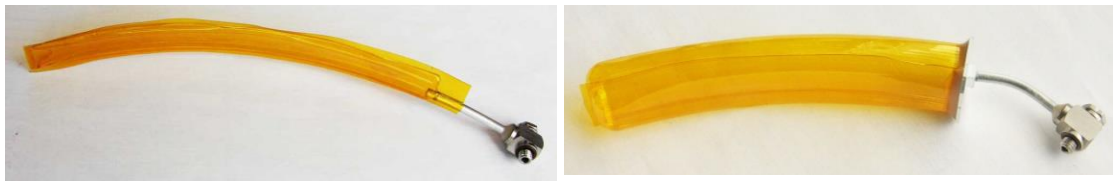

**Figure 11.  (Left) Typical collimator and (Right) camera couplant expansion bags.**

The camera expansion bags required a significantly larger volume than the collimator bags, and the design of the bag was changed to make thermoforming possible. The Aurum was formed over a mandrel and heat sealed along three edges. After cooling the mandrel was removed leaving a hole in which the end cap and tube were bonded.  The NaCl lens required the largest expansion bags.  Following several unsuccessful attempts to fabricate the large bags for this group we decided to use two bags connected with a Y-fitting.   Figure 12 shows the salt lens bezel and the double expansion bag.  We found that that the expansion bags failed when pressurized to 0.083 MPa, while in operation the bags typically see 0.0034 MPa with a 1 g couplant pressure head.

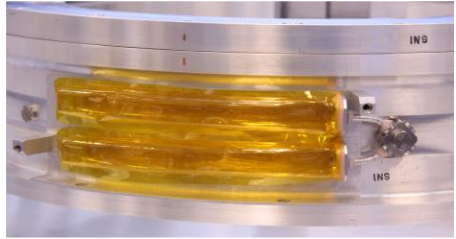

**Figure 12.  Double couplant expansion bag for the NaCl lens.**

Once the seals were installed the lenses were assembled into their respective multiplets with Viton O-rings providing a seal between the bezels.  Figure 13 shows a multiplet being filled with couplant in a fixture that rotates in two axes. The lens is positioned in multiple orientations to release any trapped air. The final air bubbles were extracted using suction from a syringe with a 0.3 mm Teflon tube inserted through the fill hole.

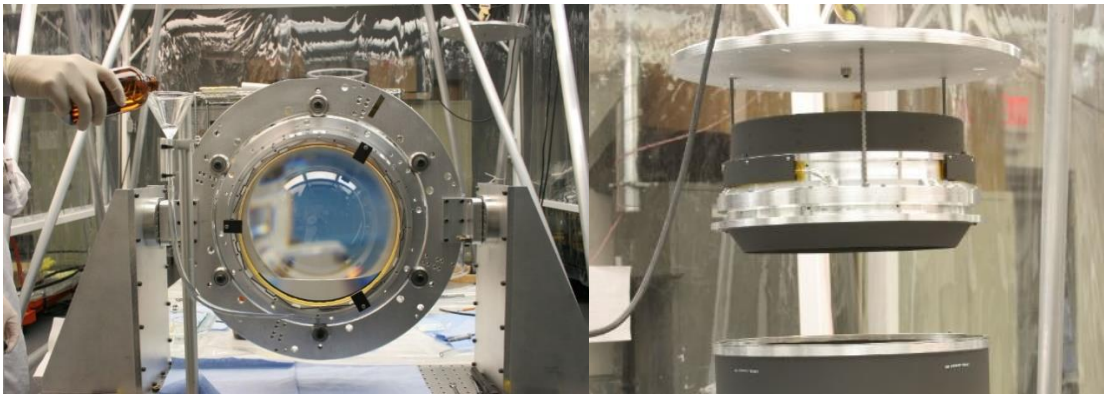

**Figure 13.  (Left) Lens group filled with couplant. (Right) A completed group with expansion bags installed being lowered into its supporting structure.**

## 4. Mechanical Design

### *4.1 Structure*

Binospec's total mass of three tons presses the upper limit of allowable instrument mass for the MMT, so designing a stiff and efficient structure was a priority.  Binospec's structure (Figure 14) has four main components: 1) an optical bench that is the instrument's backbone, 2) the mounting flange, 3) an A-frame truss connecting the optical bench to the mounting flange, and 4) a secondary bench and truss that supports the slit mask assembly, the guiders, and the periscope fold mirrors.  Table 4 summarizes the masses of Binospec components.

**Table 4.  Binospec mass breakdown**

| Component | Mass (kg) |
|---|---|
| Structure | 903 |

| Optics and mounts | 772 |
|---|---|
| Mechanisms | 593 |
| Electronics | 250 |
| Thermal shroud | 135 |
| Calibration screen | 87 |
| Total | 2740 |

The mounting flange is 400 series stainless steel for corrosion resistance and for thermal expansion compatibility with the carbon steel telescope rotator flange. The main A-frame truss members are 300 series stainless steel for low thermal conductivity between the mounting flange and the optical bench. The rest of the structure is aluminum to minimize Binospec's weight. Three A-frame trusses support the secondary bench from the main optical bench. The optical bench is a 2.3 m diameter, 0.2 m thick, closed-back honeycomb structure. All the optical components except the periscope fold mirrors mount to the bottom of the optical bench (Figure 15 and Figure 16).

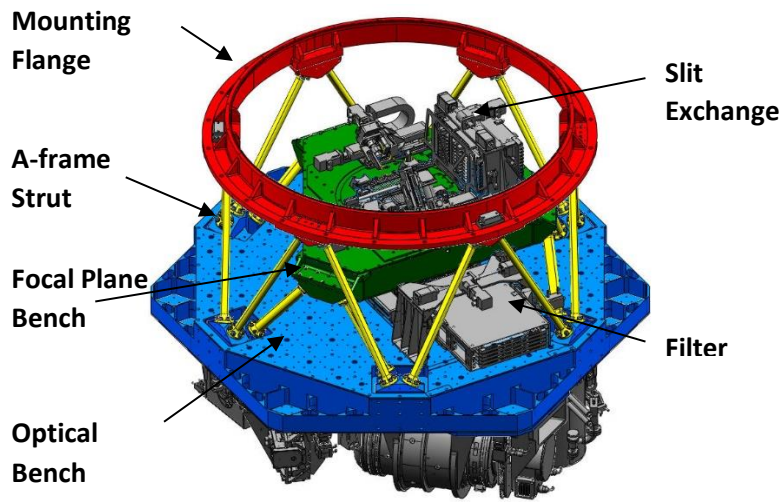

**Figure 14. A CAD model of the Binospec structure with the covers removed.**

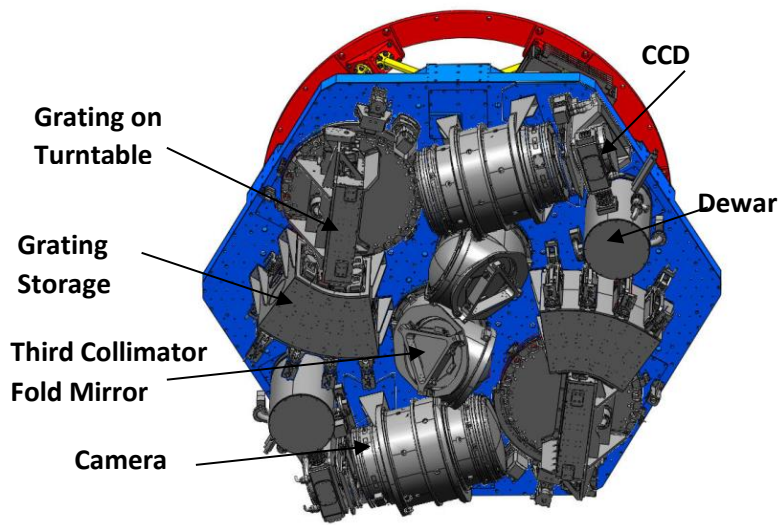

**Figure 15. A CAD model of Binospec from below with the covers removed**

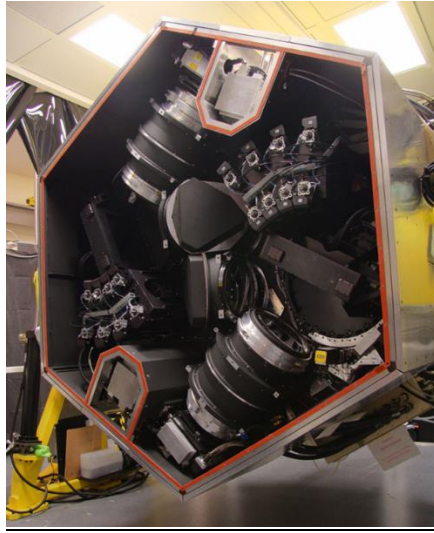

**Figure 16. Photograph of the underside of the optical bench with the bottom cover removed.**

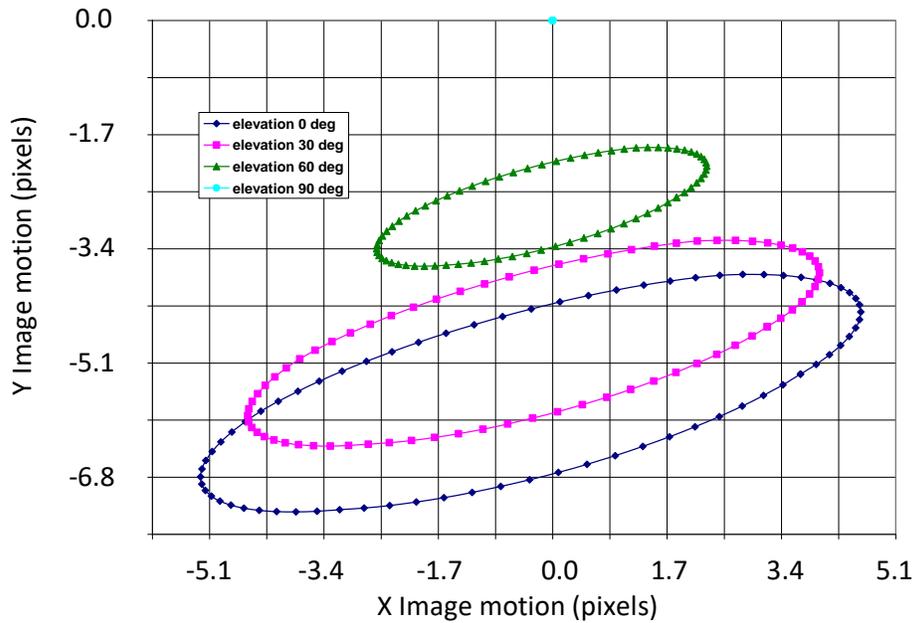

**Figure 17. Finite element/ray tracing estimate of image motion at the detector due to gravity acting on the complete instrument. Each pixel is 15μm by 15μm. The different points at each elevation represent different instrument rotator angles.**

We estimated the deflections of the fully loaded structure with finite element models.  As the telescope is tipped from zenith to horizon the lateral deflection of the optical bench relative to the MMT's instrument rotator is ~110 μm and the center of the optical bench sags ~45 μm relative to the attachment points of the A-frame struts.  The maximum relative deflection of the inner focal plane bench and the main optical bench is ~1 μm under axial gravity loads, and less than 15 μm under lateral gravity loads.   The maximum stress of 37 MPa in the A-frame struts at the lowest operating temperature with a 3g acceleration is <15% of the yield strength.

We carried out finite element modelling of the complete instrument and used optical ray tracing to predict the image motion and defocus at the detector due to gravitational flexure as Binospec is tilted and rotated at various orientations relative to the zenith. Figure 17 shows the predicted image motion at the detector due to gravity acting on the complete instrument, including the lens assemblies, fold mirrors, gratings, structure and mechanisms at telescope elevation angles of 0°, 30°, 60°, and 90º. Figure 18 shows the predicted focus shifts due to gravity acting on the complete instrument.

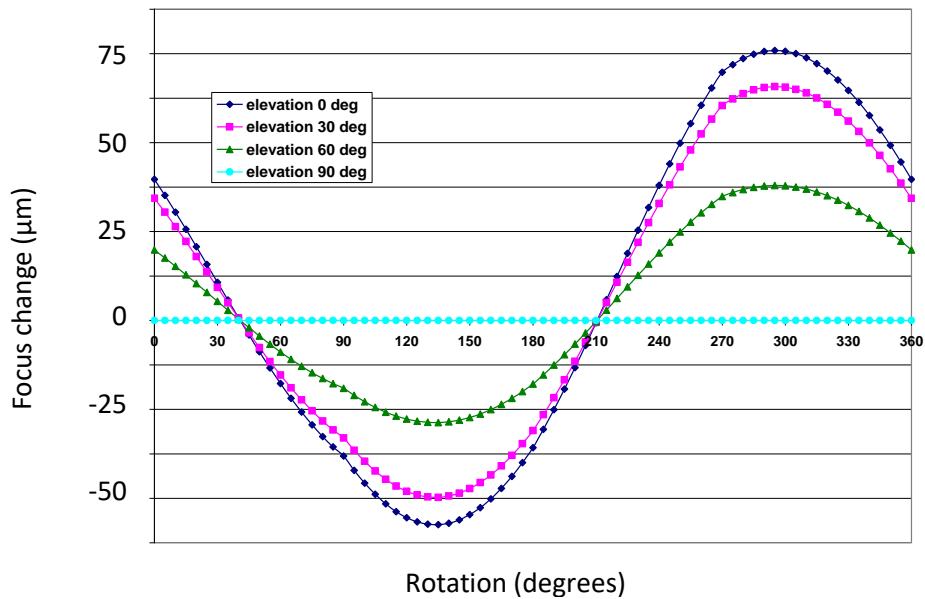

**Figure 18. Finite element/ray tracing estimate of focus change at the detector due to gravity acting on the complete instrument.**

We measured the actual flexure in both of Binospec beams at three zenith angles (0°, 30°, and 60°) and two rotation angles (the rotation angle shown in Figure 16 and rotated 90° about the optical axis) with Binospec mounted on the telescope simulator. The maximum flexure observed with the three gratings and the mirror inserted was 6.8 pixels, in reasonable agreement with Figure 17 that predicts a maximum flexure of 7.7 pixels at a zenith angle of 60°. The maximum focus change was 12 μm, considerably less than the predictions in Figure 18. We have not attempted to resolve the modelling error, but the measured focus stability is excellent. In operation, this flexure and defocus is removed by the flexure control system.

*4.2 Mechanism Overview*

Binospec's main mechanisms are a slit mask exchanger, a filter exchanger, and a grating exchanger.  The observer can choose between ten slit masks, six filters and four gratings. The control electronics are mounted in external boxes carried by the instrument.

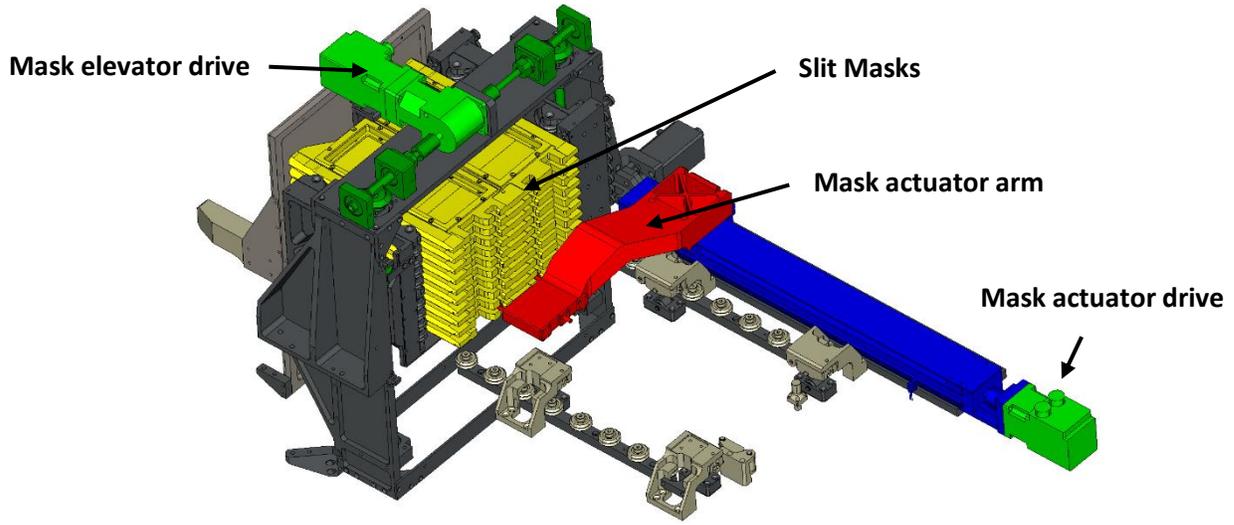

**Figure 19.  The slit mask exchange mechanism with the mask arm aligned with the lowest mask in the mask elevator.  With the mask arm deployed in this position, the elevator can be moved vertically to select any of the ten installed masks.**

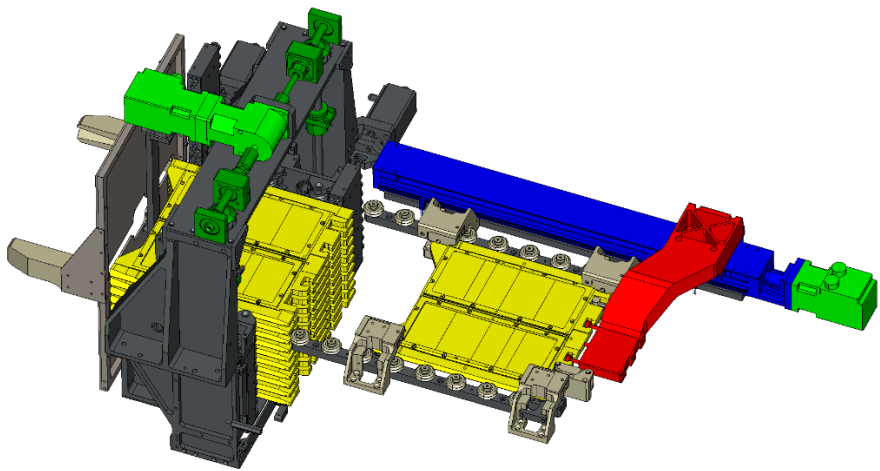

**Figure 20.  The slit mask exchange mechanism with the topmost mask selected and pulled into position at the telescope's focus.  The mask arm rides on a linear stage that can pull the mask into the focal position or reinsert the mask into the slit mask elevator.  When pulled into the focal position, preloaded rollers locate the mask in the vertical direction and the arm pulls the mask against hard stops.  Springs in the mask arm fingers apply a preload against the hard stops.**

The slit mask and filter mechanisms consist of a cassette assembly on a vertical stage to select a slit mask or filter and an actuator assembly used to move the selected item between storage and operating positions (Figure 19 and Figure 20). The lightweight cassettes provide compact storage for these interchangeable parts. Variants of this design have been used in a number of previous instruments.  Figure 21 is a photograph of the filter changer mechanism mounted on the main optical bench.

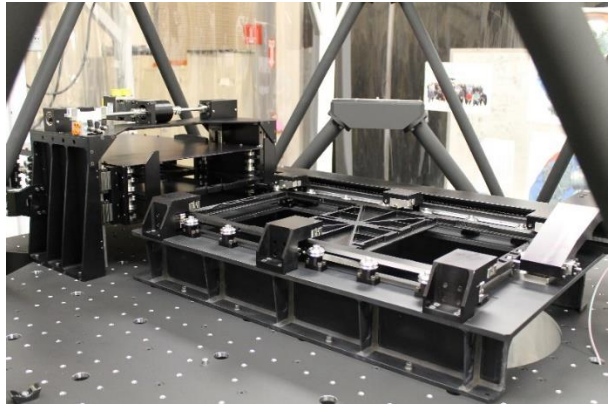

**Figure 21.  Filter exchange mechanism**

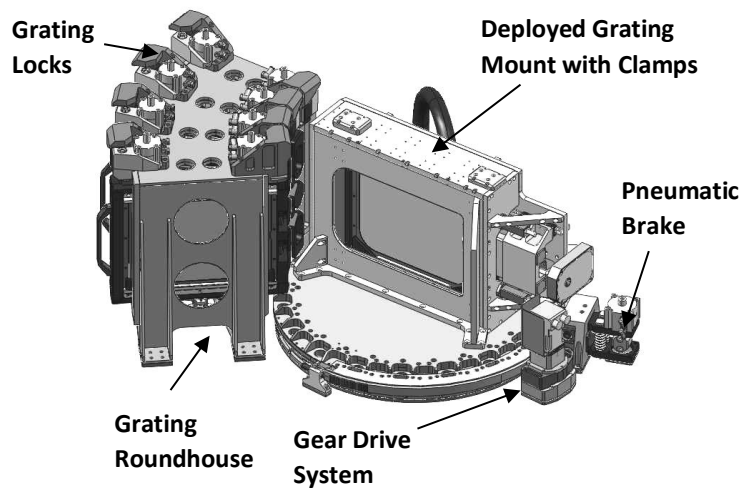

**Figure 22. Grating exchange/grating rotation mechanism**

Binospec accommodates six filters: normally g, r, i, and z, and two long pass filters for spectroscopy. Any of the filters can be replaced with alternate user-supplied filters. Pairs of filters for Binospec's two beams are mounted in a common frame to save weight, space, and complexity.

The grating exchange mechanism differs from the slit and filter mechanisms (Figure 22).  We use a rotary turntable with fixed positions for each stored grating, much like a locomotive roundhouse. The turntable provides the motion required to align gratings with the fixed storage positions as well as to set the precision grating tilt during operation, eliminating the need for a separate cassette stage.

Binospec can accommodate three gratings and a fold mirror for imaging in each of its two beams.  Tips and tilts of the grating mount and its rotary stage are potentially the biggest contributors to flexure-induced image motion, with an image motion sensitivity of 4μm arcsec$^{-1}$.

Each grating mounts in a bezel attached to the rotary grating turntable when in use. The turntable rotates on eighteen discrete THK blocks and segmented curved rails that provide an overturning moment stiffness of $10^8$ Nm/rad. This approach provides superior overturning moment stiffness and significantly reduces the start-up friction associated with conventional preloaded angular contact ball or roller bearings.

The turntable is servomotor driven with a 120:1 harmonic drive using a pinion and drive-arc gear arrangement. The measured position accuracy and repeatability is better than 10″, dominated by the gear backlash.   We remove most of the backlash by approaching the desired grating angle from a consistent direction. Once the desired grating angle is set, a pneumatically activated brake clamps the turntable.

When not in use, the gratings are held in slots orientated radially from the turntable axis with two pneumatically activated pins. A servo-driven linear ball screw mechanism engages slots on the grating bezel to load or unload the grating from the turntable. The drive and clamp approach is the same as for the slit mask and filter mechanisms. In all cases the clamps provide six degrees of constraint with a 3 g preload.

Most of Binospec's subsystems are modular, including the electronic enclosures, lens mounts, the slit mask, filter, and grating mechanisms as well as the science camera and calibration systems. This approach allows each assembly to be tested and debugged prior to integration of the completed unit. For extensive repairs, these subsystems are removed from the instrument, repaired offline and reinstalled. This approach allows us to build spare assemblies that can be used to minimize downtime in the event of a failure. We have provided good access to service motors, electronics and CCD cameras so that they can be tested or serviced through access covers.

## 5. Thermal Design

We minimize internal thermal gradients by enclosing Binospec in insulated covers (Figure 23 and Figure 24). Where possible, heat sources are kept outside the thermal enclosure. Liquid coolant removes waste heat from the external electronics boxes. The insulated covers have a 50 mm extruded polystyrene foam core with an R value of ~10, providing a ~30 hour thermal time constant. Brown et al. (2002) provide a detailed thermal analysis of Binospec.

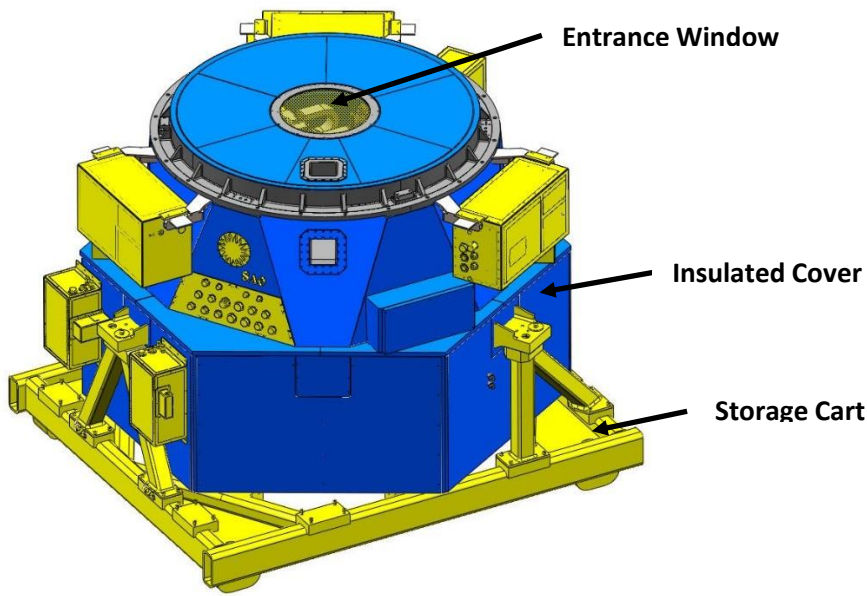

**Figure 23. CAD model of Binospec on its storage/handling cart**

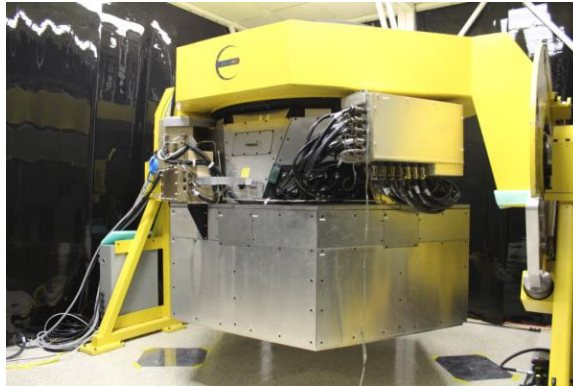

**Figure 24.  Binospec on the telescope simulator with insulated covers installed.  Access to load slit masks, filters and gratings is gained by removing the appropriate panel or cover.  A fused BK7 entrance window completes Binospec's thermal seal.**

Motors, encoders, limit switches, solenoid valves, temperature sensors, science cameras, and guide cameras must be inside the thermal enclosure. The guide cameras, encoders, dewar electronics, and the flexure control stages are powered continuously, but the motors and most feedback devices run at a low duty cycle (~2%). During normal operation the total power dissipation inside the instrument thermal enclosure is ~25 watts.  To promote convection and to help maintain a uniform temperature on both sides of and within the main optical bench, we provide ~82 25 mm diameter holes through the top and bottom face sheets of the optical bench.

The support electronics, including motor drives, controllers, power supplies, calibration light sources, guider, flexure control, and CCD electronics, are mounted externally of the main insulated volume in separate thermal enclosures. The total heat generated by these components is ~785 watts. Each electronics enclosure is insulated with 25 mm of extruded polystyrene foam with an R value of ~5. Heat is extracted using a combination of liquid-cooled cold plates and heat exchangers (with fans) mounted inside the electronics boxes. The coolant is a methanol-water mixture supplied by a NESLAB chiller.

## 6. Calibration System

A retractable calibration screen and calibration light source is mounted directly above Binospec on the telescope rotator assembly (Figure 25) to allow periodic wavelength and flatfield calibration of the detector without repointing the telescope or closing the dome. The calibration system consists of a HeAr wavelength standard and continuum lamps in an integrating sphere that illuminate a rear projection screen (Figure 26).  This subsystem also derotates the power cables, Ethernet cable, pneumatic lines and coolant lines. The cables and lines are routed through the primary mirror cell and enter the flexible e-chain through a hole in the primary mirror cell's central cone. During installation, the electrical, coolant and air connections are made at the telescope cone end and at the instrument end (Figure 27).

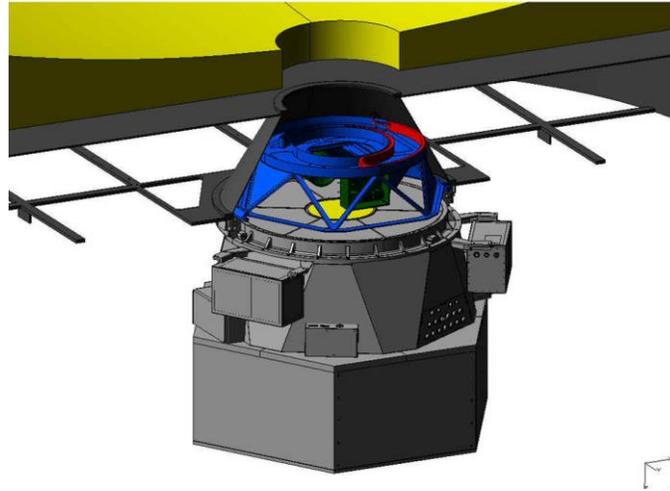

**Figure 25. Binospec and the calibration/derotator assembly shown on a cutaway view of the MMT primary mirror cell.**

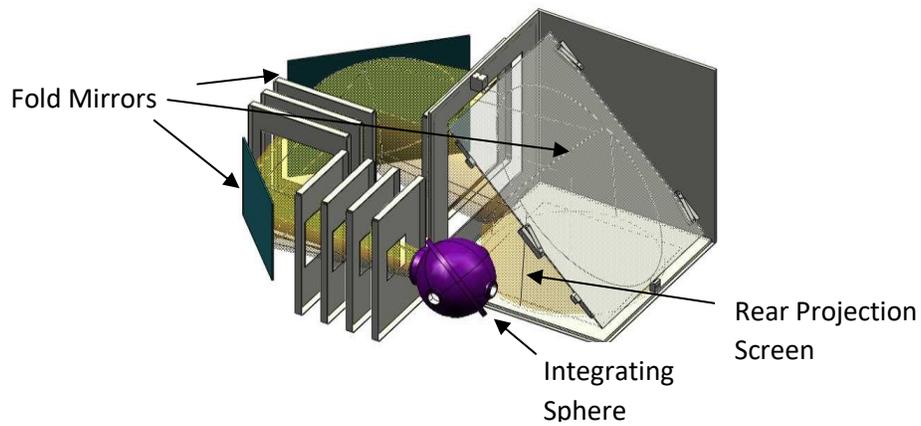

**Figure 26. Calibration system fed with light emanating from an integrating sphere (purple, at center). The light is baffled and folded three times before striking a rear projection screen.**

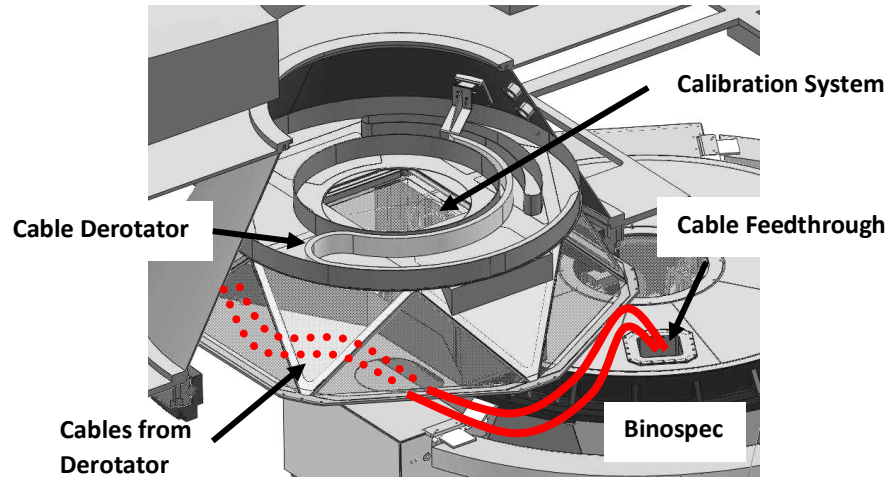

**Figure 27. Binospec along with the retractable calibration screen/cable derotator mounted on the MMT rotator bearing.**

An incandescent bulb produces insufficient blue light.  We added twelve LED's to enhance the blue continuum light and adjusted their currents to produce a flatter response.  The LED's were mounted on a custom surface mount boards (Figure 28).   Figure 29 shows the combined LED/Xenon bulb output.

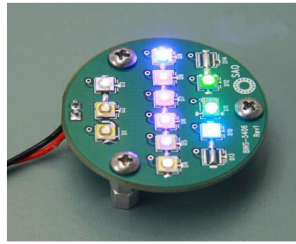

**Figure 28.   LED's mounted on custom surface board.**

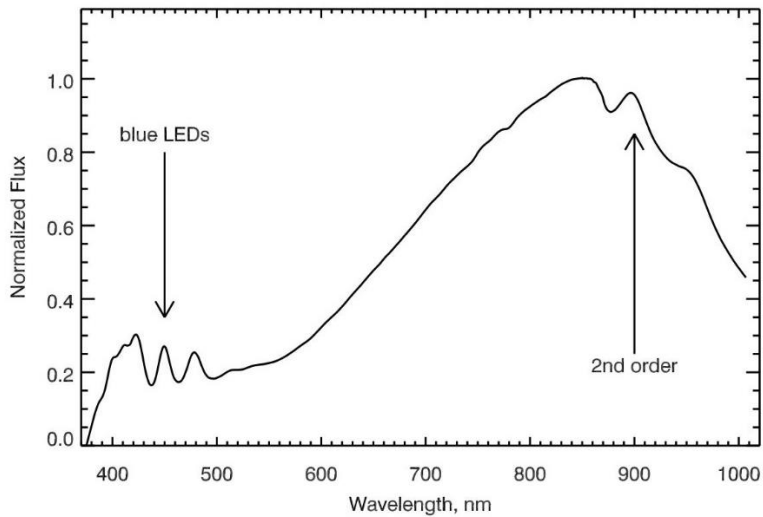

**Figure 29. Intensity plot of combined light source with LEDs and incandescent lamp.**

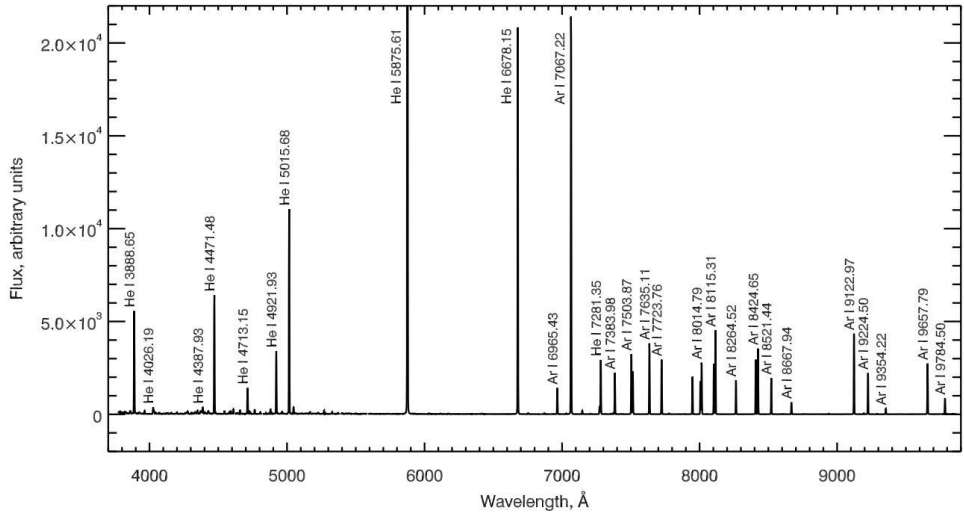

**Figure 30.  HeAr spectrum with 270 gpm grating.**

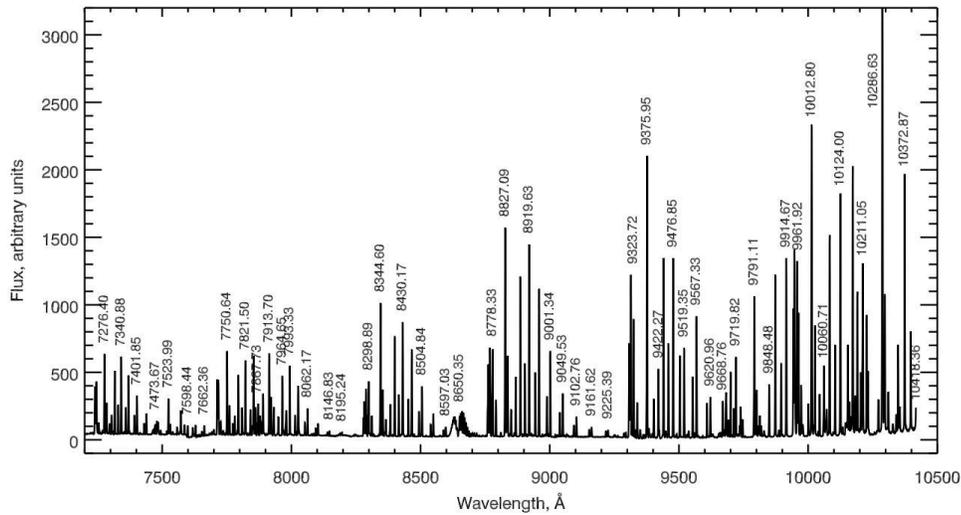

**Figure 31.  Spectrum of atmospheric emission lines with 600 gpm grating.  These lines are useful for wavelength calibration in the red.**

We use a conventional argon hollow-cathode lamp plus a Mitorika He-6201S helium lamp to produce internal wavelength calibration lines (Figure 30).

## 7. Flexure Control System

The flexure control system maintains focus and alignment perpendicular to the optical axis during a science exposure. The position and focus information are derived from images of pairs of reference optical fibers on each of two flexure control CCDs located in the same plane as each science CCD (Figure 32). One fiber is placed ahead of focus and the other behind focus, and their relative image diameters allow us to derive best focus (Figure 33). The flexure control light is injected onto the first periscope fold mirror behind Binospec's slit masks. The corrections are applied to a five-axis piezoelectrically-actuated motion stage that supports the science and flexure control CCDs in each dewar. These stages are custom Physik Instrumente parts with the motion ranges specified in Table 5. PI Nexline actuators are used for all axes to achieve sufficient travel while meeting the stiffness and load capacity requirements.

The flexure-control optical fibers are illuminated by one of two penray lamps (ArHg or Neon). Nine selectable narrow band filters (Table 6) select a single line or in some cases pairs of lines. For a given central wavelength the closest flexure control wavelength is chosen. Laboratory calibrations provide us with X and Y reference positions and fiber image diameters for best science CCD focus as a function of grating selection and central wavelength. For imaging the system works in the same fashion.

The flexure control system returns to the calibrated X and Y positions to an accuracy better than 0.1 pixel in both axes. The focus is maintained to an accuracy of 2.5 μm, limiting the image blur to a negligible 1.3 μm at the f/2 camera focal ratio. The flexure system maintains this performance over all Binospec orientations at the telescope and the full temperature range encountered at the MMT.

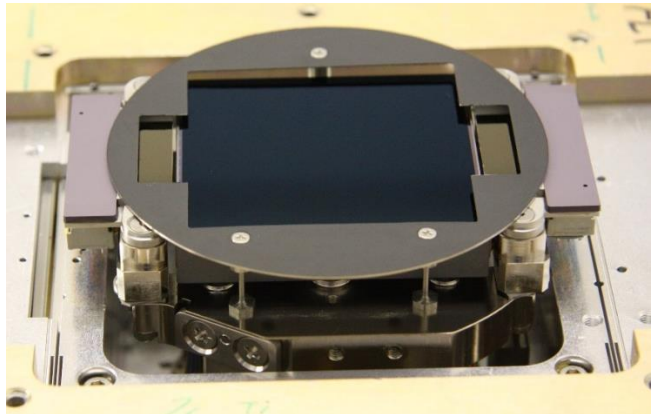

**Figure 32. Binospec science and flexure control CCDs for one of two beams.**

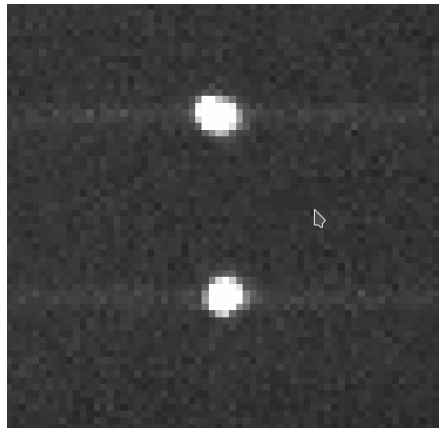

**Figure 33. Image of a pair of optical fibers from one of four flexure control CCDs.**

**Table 5.  Range of PI Flexure Control Stages**

| Motion | Range |
|--------|-------|
| X | ±500 μm |
| Y | ±500 μm |
| Z | ±750 μm |
| tip | ±2 mrad |
| tilt | ±2 mrad |

**Table 6.  Flexure Control Wavelengths**

| Wavelength | Penray Lamp |
|------------|-------------|
| 435.84 | Hg |
| 546.07 | Hg |
| 594.48 | Ne |
| 650.65 | Ne |
| 696.54 | Ar |
| 763.51 | Ar |
| 842.46 | Ar |
| 914.97 | Ne |

## 8. Guiding and Wavefront Sensing

### 8.1 Through the Mask Guiders

**Normal guiding with slit masks is accomplished by viewing guide stars with two cooled CCD guide cameras (**Figure 34**) through 6-10″ apertures located at opposite ends of the slit masks (**

Figure 35 and Figure 58).  The guiders scan ±94mm at opposite ends of the slit mask.  A third guide camera does double duty: a movable mirror allows the camera to view the center of Binospec's field of view using a small pickoff mirror positioned on the optical axis of the telescope, or a similar small mirror mounted on a special long slit mask. However, to date all long slit observations have used long slit masks with no guide mirror, and have been aligned using guide stars viewed with the through-the-mask guiders. All three guide cameras have fields of view of 80″ by 80″.

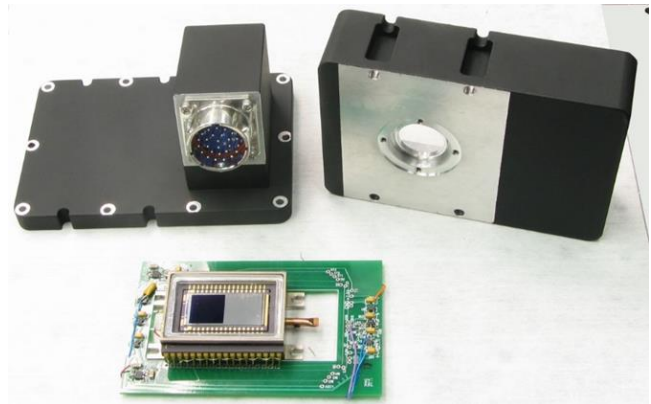

**Figure 34. Guider housing and CCD circuit board with thermoelectric cooler.**

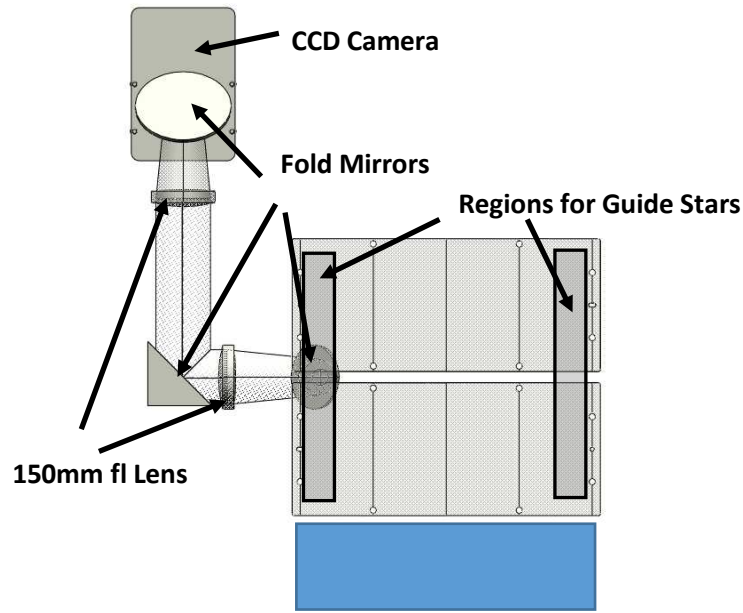

**Figure 35. Layout of through-the-mask guiders. A second set of guider optics and CCD camera patrol the right hand guider region. The wavefront sensor patrol region is shown in blue.**

Continuous wave front sensing is carried out with a dedicated Shack-Hartmann instrument that can view a strip of 5′ by 16′ at the edge of the slit mask. (Figure 35). The wavefront sensor data are used to correct the MMT's collimation and primary mirror support forces. The Binospec guide cameras are based on Carnegie/Magellan guiders (Burley et al. 2004) but with different packaging. The overall dimensions of the camera housing, exclusive of connectors, are 81mm (wide) by 46mm (thick) by 119mm (long). The Binospec guider uses an updated version of the Carnegie electronics boards modified with an Ethernet interface.

The guider CCD is the E2V CCD47-20, a 2048 by 1024 CCD with 13 μm pixels, arranged in a 1024 by 1024 frame transfer geometry. The active area of the device is 13.3 by 13.3 mm. At the focal plane scale of the MMT, 0.167 mm arcsec$^{-1}$, this format corresponds to 80″ by 80″. The thermoelectric cooler is heat sunk to the back case of the guider. The face of the camera case is then heat sunk to the instrument structure. The power dissipated is approximately 1 watt per camera.

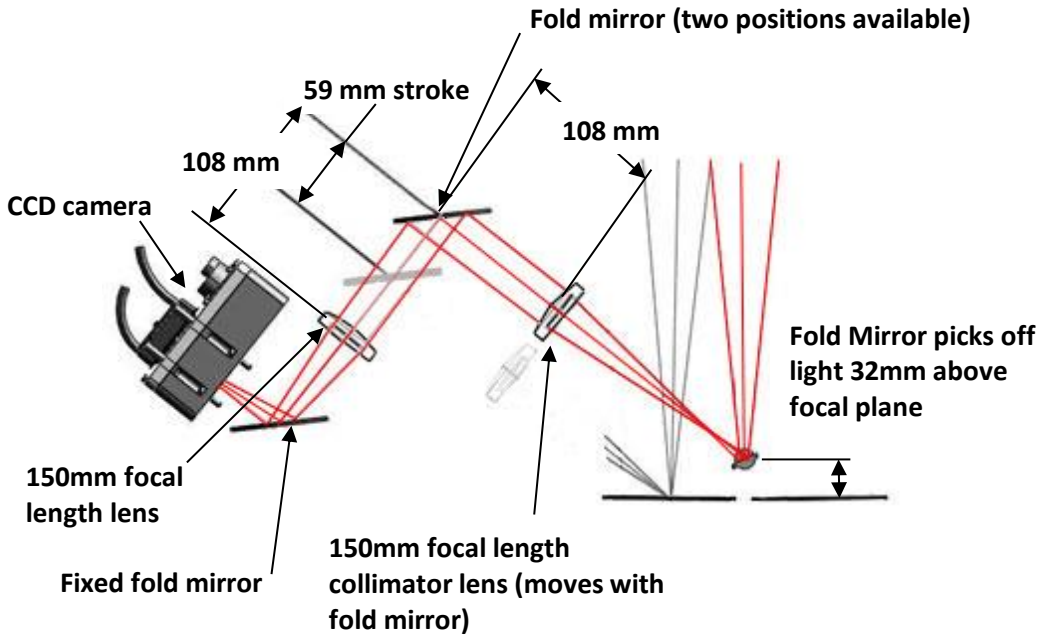

**Figure 36. Optical layout of the single object guider. A fold mirror can be moved to view an ~1′ field of view centered on-axis or on one of the slit mask halves.**

*8.2 Single Object Guider/Acquisition Camera Assembly*

The single object guider/acquisition camera is mounted on the focal plane bench slightly above and to the side of the slit mask. It allows two operational modes: central field target acquisition for pointing checks and blind offsets, and single object slit guiding. In acquisition mode, the relay lens assembly is aimed at the pick-off mirror located between the two slit masks, 32mm above the focal plane. For single object slit guiding, the relay lens assembly is repositioned to aim at the center of a slit mask with a single slit. The relay lens assembly sends the light to the camera head assembly which does not move; the relay lens/focus/fold mirror assembly is mounted on a single-axis linear stage. The optical layout of the single object guider is shown in Figure 36.

*8.3 Wave Front Sensor Assembly*

The wave front sensor is mounted above the slit mask opposite the single object guider/acquisition camera. It provides continuous wave front sensing at the edge of the slit mask over a rectangular field of 50mm by 165mm, or about 5′ by 16′, centered 139mm (~14′) off-axis. The system includes a 2-axis positioning system with a pickoff mirror. The fold mirror and focus stage are mounted on two servo-driven THK linear stages arranged in an X-Y configuration. Unlike the through-the-slit guiders and the acquisition camera, the wave front sensor requires a moving camera. For calibration, a 20 μm aperture backlit by an amber LED is mounted on a fixed stalk at a position which can be accessed by the moving head of the wave front sensor. The optical layout of the wave front sensor is shown in Figure 37.

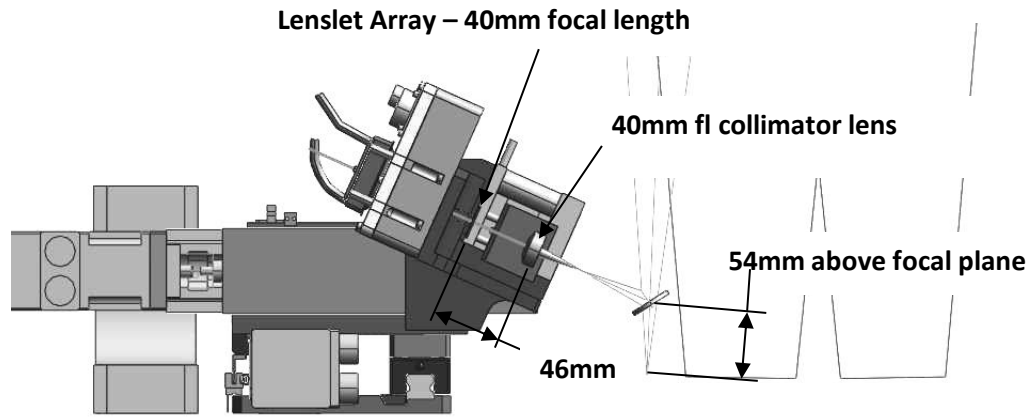

**Figure 37. Wave front sensor optical layout (side view).**

## 9. Detectors and Dewars

*9.1 CCD Complement*

The Binospec science, flexure control, and guider CCDs are all e2V devices with the characteristics summarized in Table 7. The camera electronics were developed and built at the CfA, and all have direct ethernet interfaces.

**Table 7. Binospec CCD specifications**

|  | **Science CCDs** | **Flexure Control CCDs** | **Guider CCDs** |
|---|---|---|---|
| **Device ID** | CCD231-84 | CCD42-10 | CCD47-20 |
| **Substrate** | Bulk Silicon | Epitaxial Silicon | Epitaxial Silicon |
| **Illumination** | Backside | Backside | Backside |
| **Operating Mode** | Non-inverted | Non-inverted | Non-inverted |
| **Imaging Rows** | 4112 | 512 | 1024 |
| **Imaging Columns** | 4096 | 2048 | 1024 |
| **Pixel Height** | 15.0 | 13.5 | 13.0 |
| **Pixel Width** | 15.0 | 13.5 | 13.0 |
| **Total Imaging Area** | 61.4 mm x 61.7 mm | 27.6 mm x 6.9 mm | 13.3 mm x 13.3 mm |
| **Parallel Clock Phases** | 4 | 3 | 3 |
| **Serial Clock Phases** | 3 | 3 | 3 |
| **Readout Architecture** | Full frame | Full frame | Frame transfer |
| **Fill Factor** | 100% | 100% | 100% |
| **Full Well (typ)** | 300 ke- | 150 ke- | 100 ke- |
| **Readout Mode** | Full frame | Full frame/subarray | Full frame/subarray |
| **Pixel Readout Rate** | 100 kpix/sec | 250 kpix/sec | 250 kpix/sec |
| **Total Amplifiers** | 4 | 1 | 2 |
| **Amplifiers Used** | 4 | 1 | 1 |
| **Amplifier Sensitivity** | 7 uV/e- | 4.5 uV/e- | 4.5 uV/e- |

*9.2 Science Camera Assembly*

The science camera assembly has two major components, a detector assembly and an $LN_2$ dewar assembly, joined by a copper cold strap within a vacuum bellows (Figure 38 and Figure 39). The $LN_2$ dewar assembly is based on the ND-10 Universal Cryogenics design with a capacity of 9 l, providing a hold time of ~40 hours. Figure 40 shows the cross section view of the inner fill and vent tubes of the dewar. The internal fill and vent tubes are biased upward ~2/3 of the height to maximize the hold time while observing. The dewar is designed to operate in all attitude orientations of the telescope.

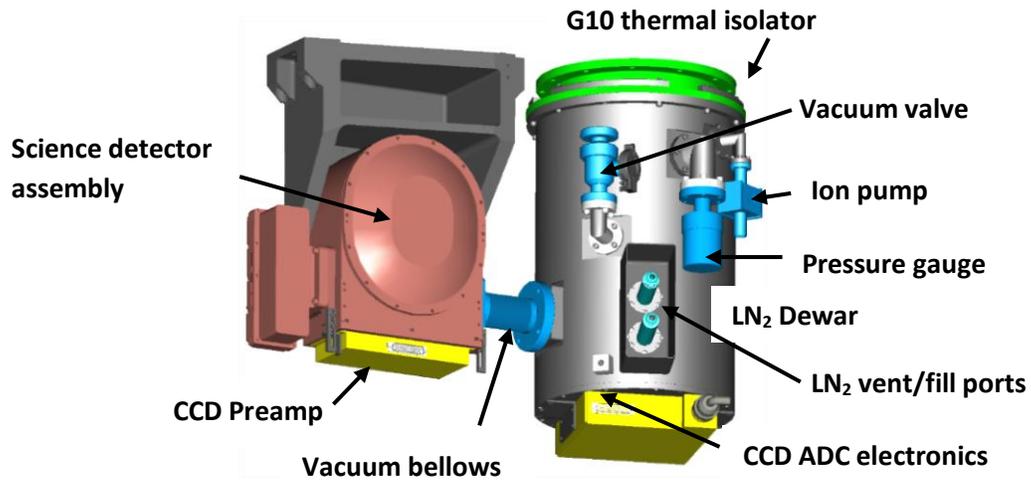

**Figure 38.** The science camera assembly (rear view)

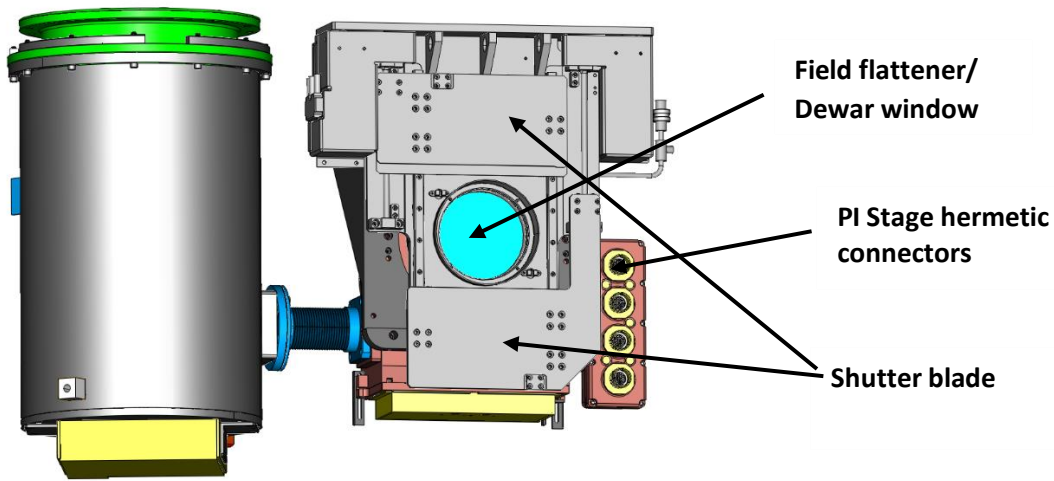

**Figure 39.** The science camera assembly with shutter (front view). The two shutter blades are individually controlled to allow scanning a narrow aperture across the detector for short exposures.

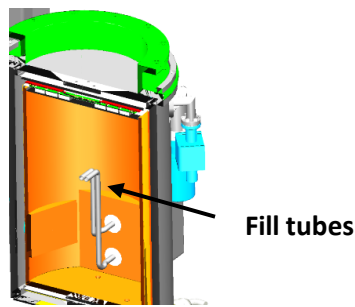

**Figure 40.** Cross section view of the LN2 dewar assembly showing the fill tube location

The dewar has a charcoal getter containing ~120 ml of activated charcoal (**Error! Reference source not found.**). When cooled to LN$_2$ temperature, the charcoal getter holds the dewar at ~5 x 10$^{-7}$ Torr.

The science and flexure control CCDs mounted on the flexure control system's five-axis Physik Instrumente stage. We use the tip/tilt control to align the CCD surface exactly perpendicular to the optical axis.

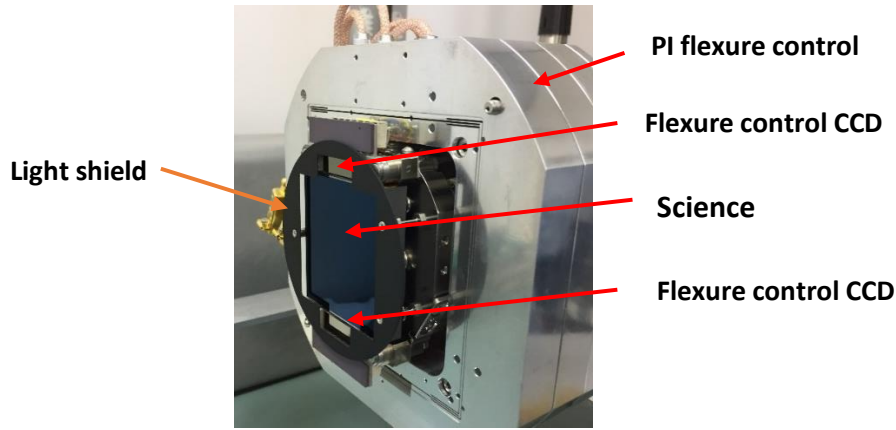

**Figure 41. The CCD array installed onto the PI stage. The science CCD and the flexure control CCDs are mounted on a common Invar cold plate, and are coplanar to 0.03 mm.**

The field flattener lens for the camera optics serves as the dewar vacuum window (Figure 42). The lens is bonded to the bezel using a continuous GE Silicones RTV60 ring 6.35 mm wide and 12.7 mm deep around the lens perimeter. A separate Viton O-ring provides the vacuum seal.

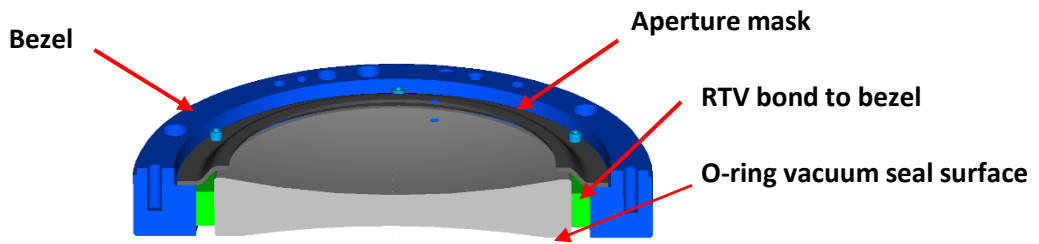

**Figure 42. Cross section view of the field flattener lens assembly.**

## 10.0 Electronics and Control

*10.1 Introduction*

Figure 43 is a functional block diagram of the Binospec electronics. Due to packaging and space constraints these functions do not map exactly to the physical block diagram (Figure 44). In the sections below we describe each of electronics functions in turn and how these functions map to external electronics boxes. Each external electronics box (Figure 45) has a circuit breaker, a Cat5e Ethernet interface, as well as temperature and power monitoring. The boxes are insulated with 25 mm of foam and are liquid cooled with a liquid to air heat exchanger and fan. The boxes contain a total of 13 Advantech ADAM-6000 series modules for Ethernet controlled I/O (including monitoring and power control functions) and three ADAM-4570 modules for RS-232/Ethernet conversion. External cabling between boxes generally uses rugged Mil-type circular connectors.

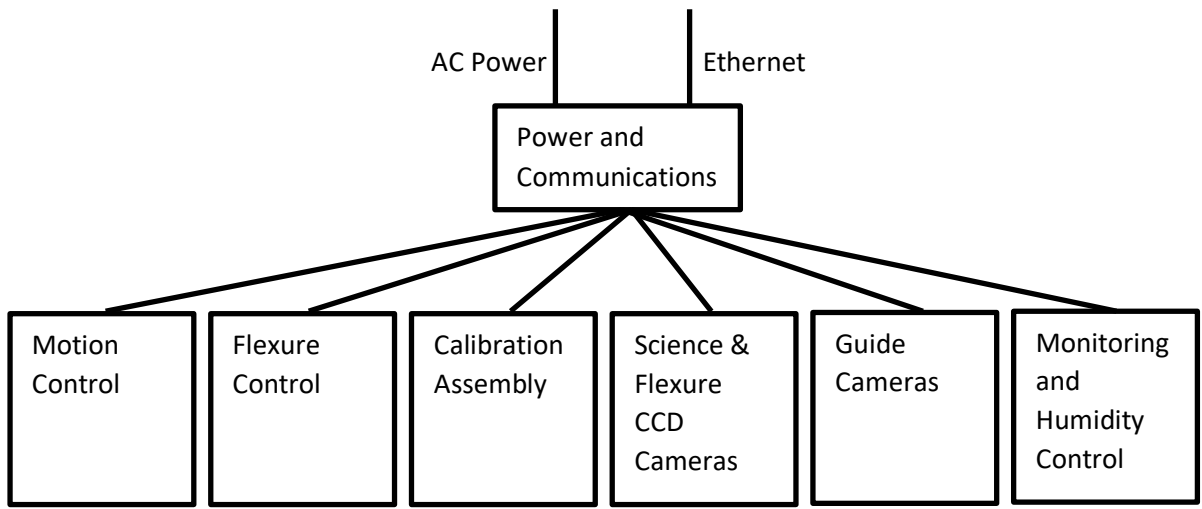

**Figure 43. Electronics functional block diagram**

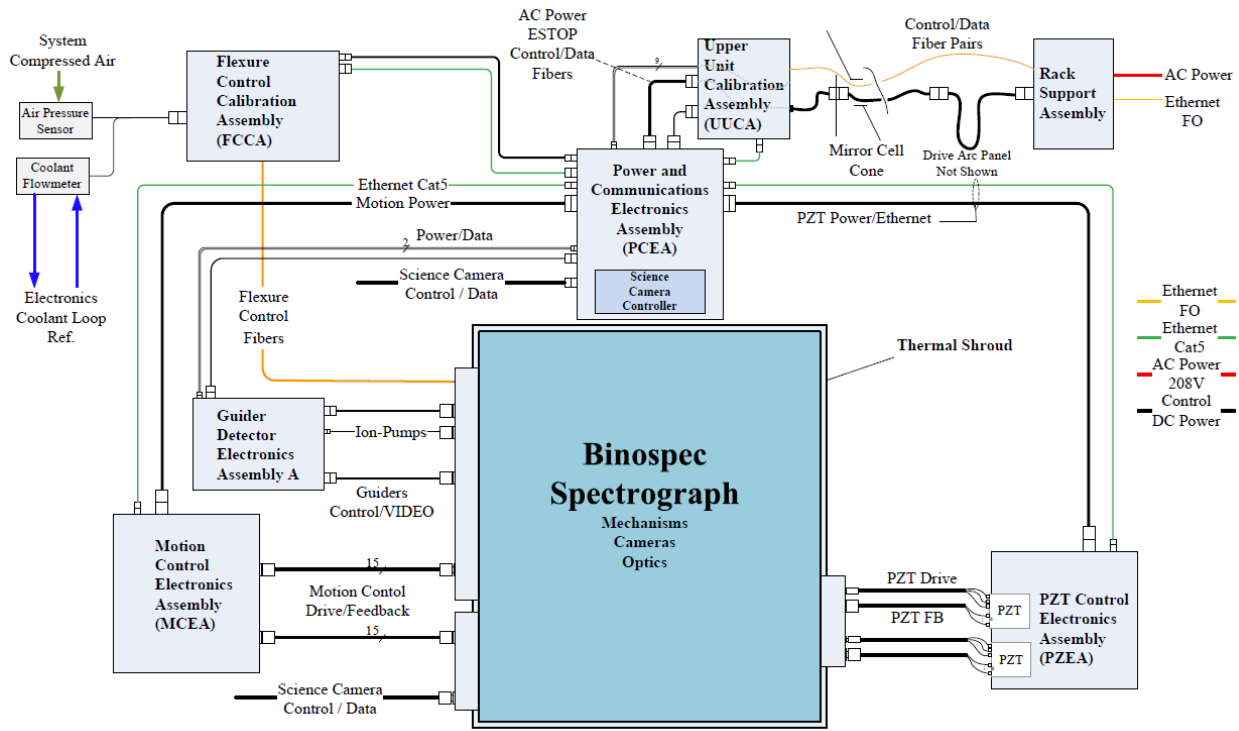

**Figure 44.** Electronics physical block diagram. Binospec's electronics are contained in six boxes external to the main instrument volume to minimize internal heat dissipation. The electronic boxes are insulated and liquid cooled to minimize heat escape into the telescope enclosure.

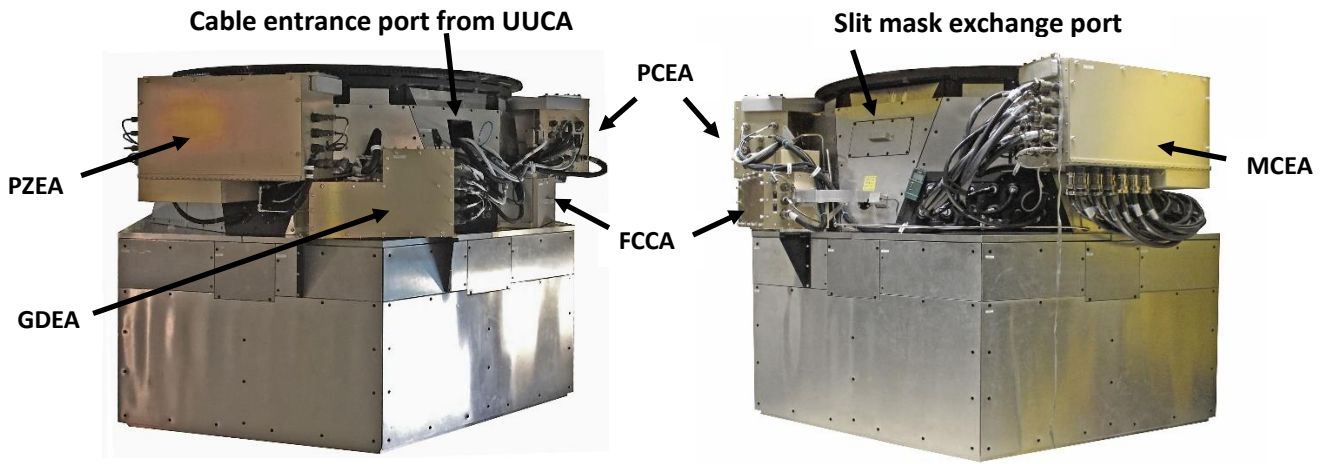

**Figure 45.** Binospec's external electronics boxes. The names are defined in the text.

*10.2 Power and Communications*

The Power Control Electrical Assembly (PCEA) accepts facility power and a duplex fiber optic Gigabit Ethernet. The input AC power is filtered, surge protected, and current limited with circuit breakers. AC power and 24VDC power is supplied to the other electronics boxes, controlled and monitored with two ADAM-6000 series modules. A portion of the digital science and flexure CCD electronics common to both beams is also located in the PCEA.

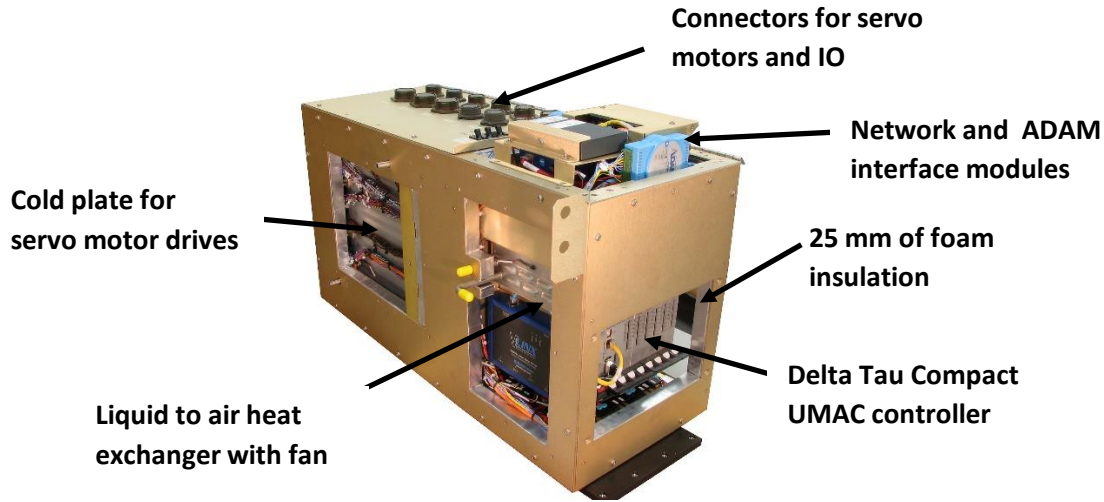

**Figure 46.** Motion Control Electronics Assembly (MCEA) with covers removed

*10.3 Motion Control*

Binospec contains 22 motion stages actuated by servo motors with encoder feedback, 10 axes of piezoelectric actuators with capacitive encoders, and 14 pneumatic actuators. The 22 motion stages include 20 linear stages and two rotary stages for the grating tilt adjustment. One of the motion stages inserts or retracts the backlit calibration screen in the calibration system (Upper Unit Calibration Assembly or UUCA) as described in section 6. This stage is controlled by a Copley Controls Accelnet intelligent servo motor drive through an RS232 interface, and an RS232 to ethernet converter. The remaining 21 axes, located within the main Binospec enclosure are controlled by a Delta Tau compact UMAC controller. The Delta Tau UMAC, 24 Copley Controls servo motor drives, and related support electronics are located in the Motion Control Electronics Assembly (MCEA) shown in Figure 46. The MCEA uses three ADAM-6000 series modules for control and monitoring, and an ADAM-4570 module for serial communications with the encoders.

*10.4 Flexure Control Electronics*

The electronics for the flexure control system are distributed between two external electronics boxes, the Flexure Control Calibration Assembly (FCCA) and the Piezoelectric Electronics Assembly (PZEA). The FCCA contains an integrating sphere, argon-mercury and neon Penray lamps to produce the flexure control emission lines, filters to select the desired emission lines, filter wheels, a shutter, power supplies, two ADAM-6000 modules for monitoring and control, and an ADAM-4570 module for serial communications with the filter wheels. The PZEA contains two five-axis Physik Instrumente controllers for the flexure control stage and two ADAM-6000 modules for monitoring and power control.

*10.5 Calibration Assembly Electronics*

The calibration electronics are located in the calibration assembly (Section 6). They include a Copley Controls Accelnet intelligent servo motor drive, power supplies, three ADAM-6000 series modules for lamp control, power monitoring, and power control. An ADAM-4570 module is used to communicate with the servo motor drive.

*10.6 Science and Flexure Control CCD Electronics*

*10.6.1 Camera Control, CCD Timing Generation, and Data Communication*

A single CCD control unit operates both science cameras. A single Ethernet communications channel on an embedded Avnet digital control module with a Xilinx Virtex-5 FPGA handles all communications with the CCD readout electronics (Figure 47). The FPGA implements a state machine that produces the timing signals required to operate the science and flexure control CCDs as well as the synchronous timing signals required for processing and digitizing the CCD output video waveforms. The timing signals are simultaneously transmitted to the CCD clock driver and signal processing module in the two cameras. The digitized CCD pixel data from both cameras are collected in the CCD Control Logic unit and are then transferred to a host computer over the ethernet interface.

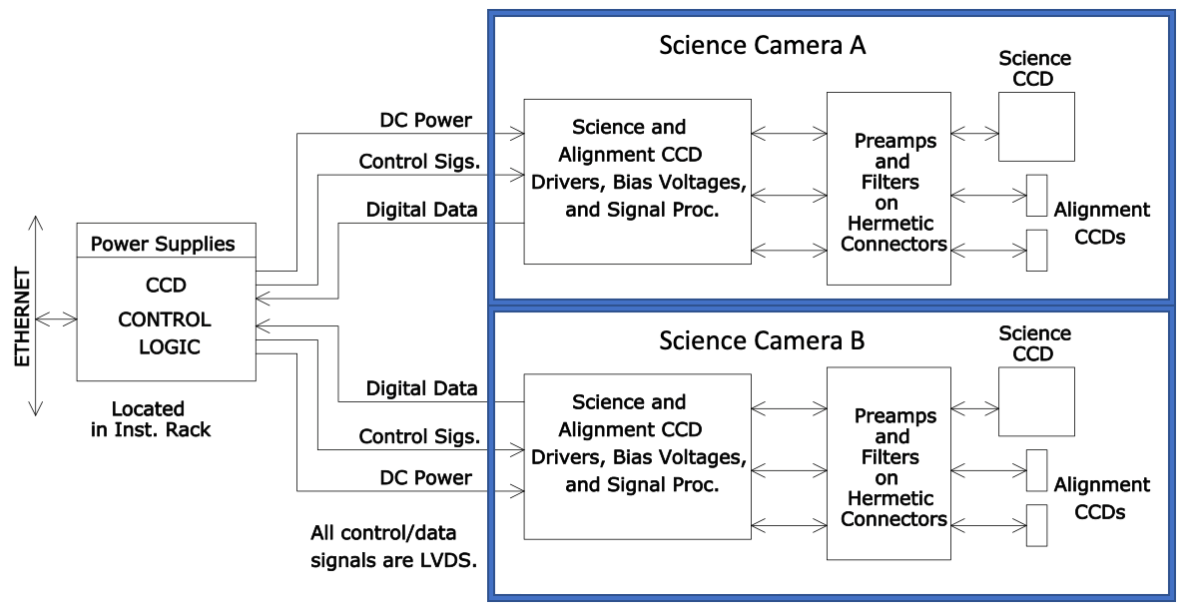

**Figure 47. The two science cameras are controlled as a single unit using a common set of timing and control logic.**

*10.6.2 CCD Readout Electronics*

For each camera a single set of electronics operates and reads out the science and flexure control CCDs. During science image integration the flexure control CCDs acquire flexure calibration images. After the science integration is complete, the electronics operate and read out the science CCDs. The readout electronics are housed in two modules per camera. The preamplifier modules supply bias voltages to the CCD output amplifiers and CCD video signal preamplifiers that amplify the CCD output prior to transmission to the second electronics module. The second CCD readout module houses the CCD clocking circuitry as well as the CCD video signal processing and digitization circuitry.

*10.6.3 CCD Clock Drivers*

All CCD clocking rail voltages are produced by DC voltage regulator ICs and operate at fixed voltages. The CCD clocking logic allows the four output science CCDs to be read out through any one amplifier, any two amplifiers, or all four amplifiers simultaneously. The timing signals produced by the FPGA are converted to the voltage and power level required for driving the science and flexure control CCDs by simple analog switches. The CCD reset gate clock signals are buffered via operational amplifiers in the preamplifier module prior to application to the CCD.

*10.6.4 DC Bias Generators and Video Preamplifiers*

The preamplifier module, mounted on the science camera vacuum housing, produces the DC bias voltages required for the CCD output amplifiers and external amplifiers for the CCD output video signal. The amplified signal is sent to the camera control module for further processing and digitization. The fixed DC bias voltages are produced with integrated circuit voltage regulators and fixed resistors.

The preamplifiers accept signals directly from the output source nodes of the CCD output amplifiers and provide a constant current load of ~3 mA. A DG419 analog switch is used as a DC restoring clamp and low noise JFET transistors buffer the signal before a low noise operational amplifier stage. The amplifier circuit includes a switchable voltage gain of either 4 or 7. The preamplifier output is transmitted to the camera control module for processing and digitization.

*10.6.5 CCD Signal Processing*

The camera control module includes a multiplexed input that selects either the science CCD or flexure control CCD signal for processing and digitization. The first processing stage is a gain stage with a selectable voltage gain of 1 or 2. The output of the first stage is applied to an integrator stage that performs correlated double sampling using a dual slope integrator. The output of the integrator is buffered by a pair of operational amplifiers, one inverting and one non-inverting, that drive the inputs of an 18-bit analog to digital converter (ADC). The Analog Devices AD7982 converter is used at 16 bit resolution. The output of the ADC is serially transmitted to the FPGA for transmission to the host over the Ethernet interface.

*10.7 Guide Camera Electronics*

The control electronics for the four guider CCD cameras (the two through the mask guiders, the single object guider, and the wave front sensor camera) are located in the Guide Detector Electronics Assembly (GDEA). The GDEA also contains ion pump controllers and vacuum gauge readouts for the two science/flexure dewars. The GDEA uses two ADAM-6000 modules for control and monitoring.

*10.8 Monitoring Electronics*

Power and temperature monitoring inside the external electronics boxes is provided by ADAM-6000 series modules interfaced to sensors as described above. Inside the main instrument enclosure temperature and humidity monitoring is provided by 1-Wire sensors interfaced to an Embedded Data Systems HA7NET Ethernet to 1-Wire adapter.

*10.9 External Electronics Rack*

Electronics in an external rack accepts 208V three phase power from the facility and provides filtered, surge protected, and current limited power to Binospec. The AC power for the CCD camera power supplies is isolated with a shielded transformer. The power is controlled through an Ethernet controlled multichannel switch. Two satellite racks are available to power Binospec in the repair/maintenance facility and in its storage bay on the observing level of the MMT.

## 11. Support Facilities

*11.1 Telescope Simulator*

The development of large instruments requires a surprising amount of lifting, handling, assembly and alignment equipment. An MMT telescope simulator played a large role in the construction and test of Binospec (Figure 48). This fixture reproduces the instrument mounting interface at the MMT and allows ±180° instrument rotation in elevation and azimuth. The instrument support ring is supported on two spherical roller bearings that are mounted on two truss supports bolted to the concrete floor of the laboratory. The elevation rotation is provided by a hydraulic piston connected to a large wheel on one side of the ring. A hole pattern in the wheel allows rotation in 10° increments, limiting the maximum freefall to 10° in the event of a hydraulic failure. In this event, the ring contacts a large spring that limits the dynamic shock response transmitted to Binospec at ~1 g to prevent damage to the optical components. A class 10000 clean tent was installed surrounding the simulator, leaving enough clearance to allow for full ±180° Binospec rotation.

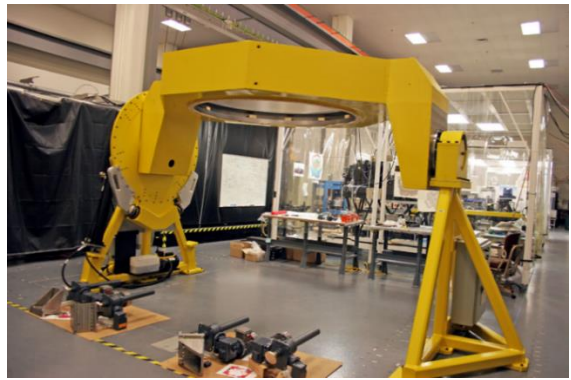

**Figure 48. MMT telescope simulator for Binospec testing.**

The major optical assemblies are mounted on the underside of the main optical bench (Figure 49). Rotating Binospec upside-down during the installation of these assemblies allowed us to safely lower and position these assemblies onto the bench with an overhead gantry crane (Figure 49).

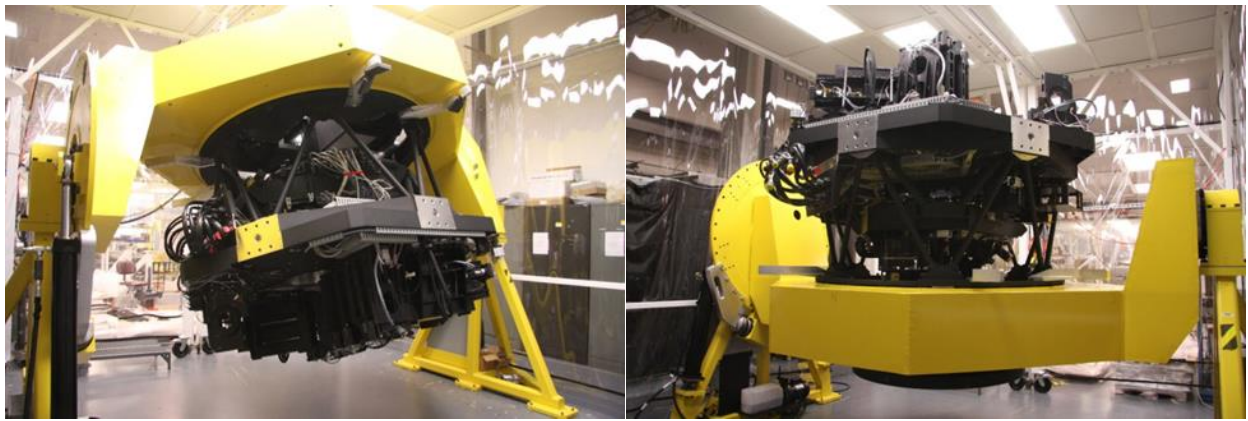

**Figure 49. (Left) Binospec tilted to a 30° zenith angle on the telescope simulator. (Right) Binospec inverted on the telescope simulator.**

For system testing and calibration, the calibration assembly was mounted to the top side of the simulator and opaque curtains were installed. Binospec was deployed over the full range of elevation and azimuth angles to verify the functions of the mechanisms and the flexure control system. The image motion without flexure control matched the finite element predictions.

*11.2 Instrument Repair Facility*

An instrument maintenance facility is located across a parking lot from the main MMT enclosure. This facility was upgraded to meet Binospec's electrical requirements and to support the large and heavy telescope simulator with its clean tent. A low profile two-ton 3-axis crane was installed to allow Binospec service with Binospec mounted on the simulator. A clean tent and laminar flow bench for delicate CCD and electronic work were installed.

*11.3 Slit Mask Cutter*

The LPKF Stencil Laser slit mask cutting machine (see Appendix 2) will reside in the basement of the Common Building below the summit. This area has been rebuilt with an environmental control system to meet the laser mask cutter's temperature and humidity requirements. The refurbished space will also be used for storage of the MMT's f/5 instrument spare parts.

## 12. Performance and Conclusions

Binospec was commissioned in November and December 2017. We observed several spectrophotometric standards to measure Binospec's throughput through a 5″ wide slit to minimize seeing and aperture losses. The results for all three gratings are shown in Figure 50; we include results for three tilt settings with the 600 gpm grating. We estimate the telescope and corrector throughput at about 73% (two aluminum reflections plus 10 glass-air surfaces), so to derive the throughput of Binospec alone one should multiply the results in Figure 50 by ~1.37. This calculation suggests that Binospec's peak throughput exceeds 40% with all three gratings, in agreement with our estimates.

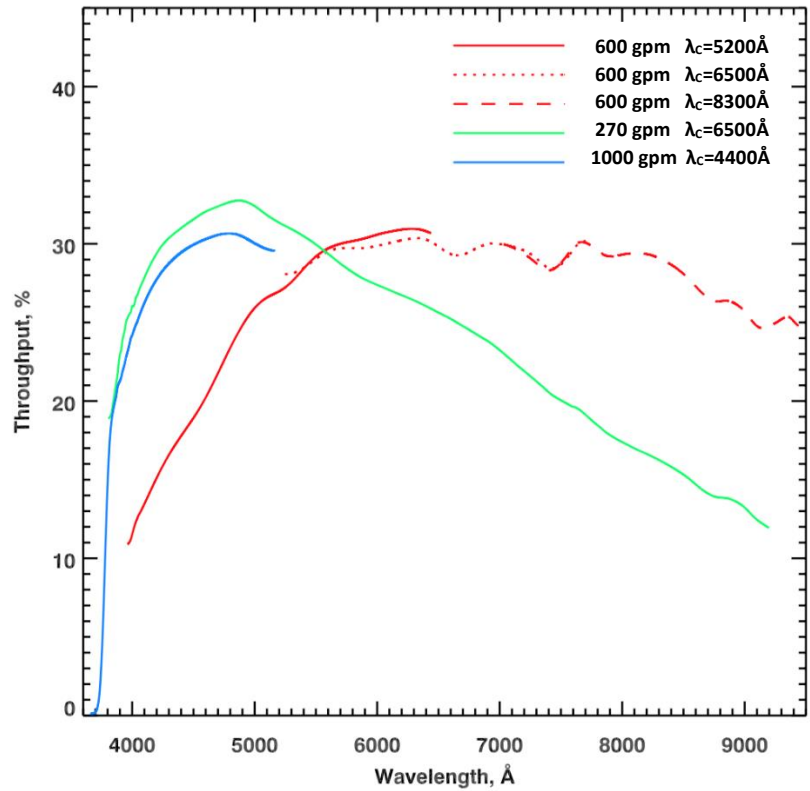

**Figure 50.** Combined Binospec and MMT throughput.

We fit afternoon observations of the solar spectrum to derive Binospec's effective spectral resolution with a 1″ wide slit. The results, which matched our expectations, are shown in Figure 51.

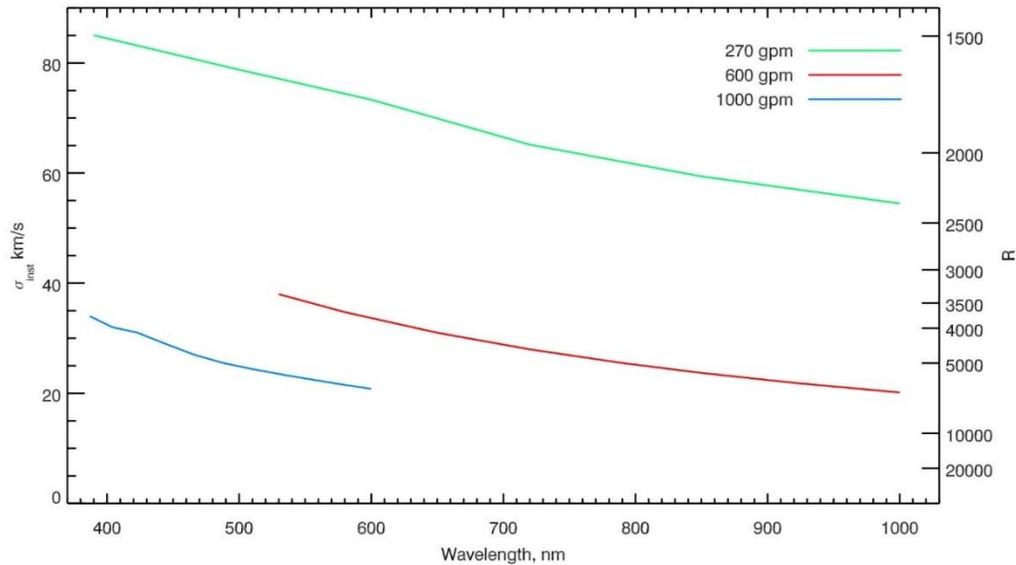

**Figure 51. Binospec line spread function (resolution) for three gratings with a 1″ wide slit measured from solar spectra.**

The read noise is 3.2 e$^{-1}$ RMS averaged over the eight signal chains on the two e2v CCD231-84-F64 CCDs, low enough to be negligible for most purposes.  Figure 52 illustrates Binospec's ability to extract high quality spectra from low surface brightness objects.   Figure 53 shows a high signal-to-noise spectrum with the blue-blazed 1000 gpm grating.

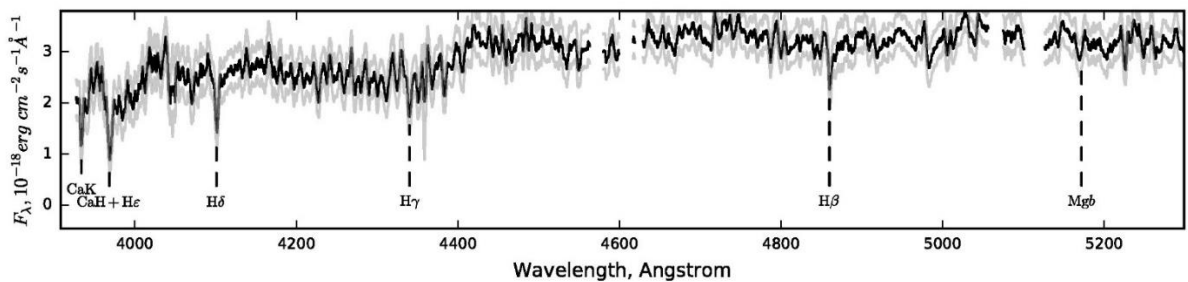

**Figure 52.  A flux calibrated spectrum of UGC5442, a dwarf spheroidal galaxy in the M81 group taken with the 1000 gpm grating and with an integration time of 3.33 hrs.  The galaxy has a surface brightness of ~25.5 mag per square arcsecond in the B band,  ~20 times fainter than the night sky between 4000-5000 Å.  The spectrum is integrated across 80 pixels (~20″), with an integrated B~22.  The data are smoothed using a 3$^{rd}$ order Savitsky-Golay  (Savitsky and Golay 1964) filter with a window size of 31 pixels. The original spectrum has a signal-to-noise ratio of 13 per pixel, allowing us to measure a stellar velocity dispersion of 12±3.5 km s$^{-1}$, about half the width of Binospec's line spread function with the 1000 gpm grating.  The unsmoothed 1σ errors are shown in light gray.**

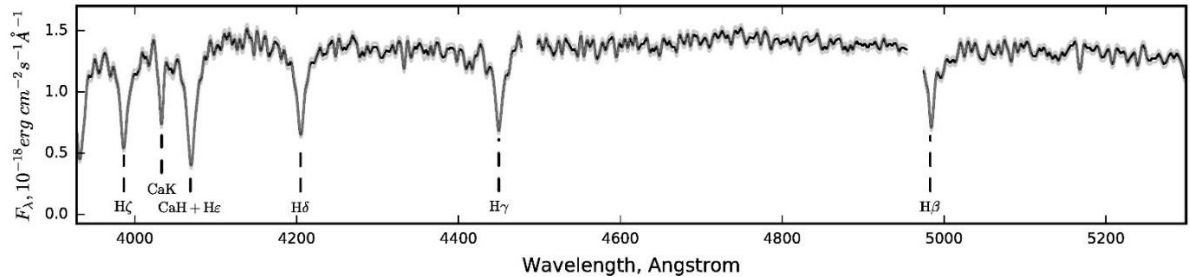

**Figure 53.** A flux calibrated spectrum of [GMP83]4348, a diffuse post-starburst galaxy in the Coma cluster, from a 2 hr integration with the 1000 gpm grating. The spectrum is from a single pixel along the slit with B~22.6, and the unsmoothed 1σ errors are shown in light gray. The signal-to-noise is much higher than in Figure 52 because the relative sky level was 80 times lower. The data are smoothed with a 3$^{rd}$ order Savitsky-Golay filter with a window size of 31 pixels.

*Facility:MMT Observatory*

**Acknowledgements**

We thank several vendors who went above and beyond to produce superior parts for Binospec: The Bechdon Company, Boston Engineering, Canon, ECI, Coastal Optics (now Jenoptik), SESO, and Tinsley.

We thank talented SAO engineers, technicians, and machinists who helped make Binospec a reality: Kevin Bennett, Roger Eng, Michael Honsa, David Weaver, and the CfA Model Shop.

We thank the ever helpful MMT Director, Grant Williams, and the MMT staff, including Will Goble, Marc Lacasse, and Ricardo Ortiz. Kirill Grishin (Moscow State University) assisted with figures in Section 12.

Funding for Binospec was provided by the Smithsonian Institution and the National Science Foundation's Telescope Systems Instrumentation Program administered by AURA.

**Appendix 1 – Materials Testing**

*Epoxy testing*

We chose Hysol 9313 for all of the structural bonds between the flexure nubs and glass. To fully characterize the Hysol 9313 epoxy, we performed several tests to measure its modulus, strength and coefficient of thermal expansion (CTE). Siltex (silica powder) is added to the epoxy (40% by weight) to increase its viscosity, allowing us to inject the epoxy for bonding without drips.

We used an Instron 3365 machine to measure the elastic modulus of Hysol 9313 using a three point bend test on epoxy beams (63.5 mm x 12.7 mm x 5 mm thick) as shown in Figure 54. We measure the modulus of the filled Hysol 9313 to be 3970 MPa at 42°C, increasing to 4275 MPa at -18°C. The modulus of the unfilled Hysol 9313 is lower: 2420 MPa at 42°C and 2820 MPa at -18°C. The tensile strength of the filled Hysol 9313 was measured to be 54 MPa at 42°C using tensile specimens with a central cross section of 5mm x 5mm. The tensile strength will increase at lower temperatures.

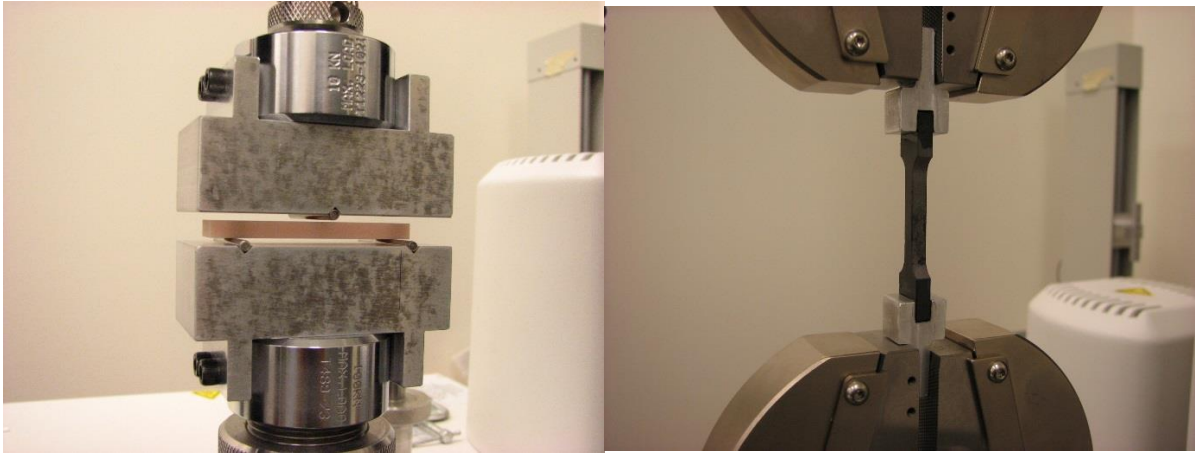

**Figure 54. (Left) Three point bend test for epoxy modulus. (Right) Tensile test to measure breaking strength**

An epoxy cylinder of 12.7 mm diameter and 25.4 mm tall was cast to measure the coefficient of thermal expansion. A thermal diode was attached to the sample for temperature measurement. The sample was chilled to -43°C and placed on a piece of G10 fiberglass sheet to insulate it from the granite test plate. A digital dial indicator is used to measure the sample dimensions as a function of temperature. The filled Hysol 9313 CTE is measured to be 52ppm/°C.

*Lens/nub strength tests*

We designed the shear test fixture shown in Figure 55 to test the glass to nub bond strength. A test sample is shown in Figure 56. The goal of this fixture is to apply a pure torque load on the specimens with negligible side loads or moments. The torque was produced by applying a vertical load with the Instron. The vertical force is reacted by the bearing mounts at each end of the shaft. The torque is applied to the glass/nub sample and reacted by a precisely supported flexure.

The results in Table 8 include all material bonding pairs. Five samples of each combination were tested. The tests were performed at 20°C and at -13°C. Samples 5, 7 and 8 were soaked in LL5610 for several weeks prior to testing. Several samples did not break. The applied force was limited to 890 N to prevent yielding the flexure at the end of the fixture. The stress levels shown for these forces with an asterisk do not represent the ultimate strength of the material but still allowed us to draw some important conclusions. The measured stress in the first sample demonstrates a safety factor of three compared with the maximum calculated stress resulting from a 3g gravity load combined with a 42°C temperature change and assembly loads. All other samples demonstrate a safety factor of five or more.

For the samples that did break the glass failed and the bonds remained intact (Figure 57). The samples that were soaked in LL5610 show no loss in strength.

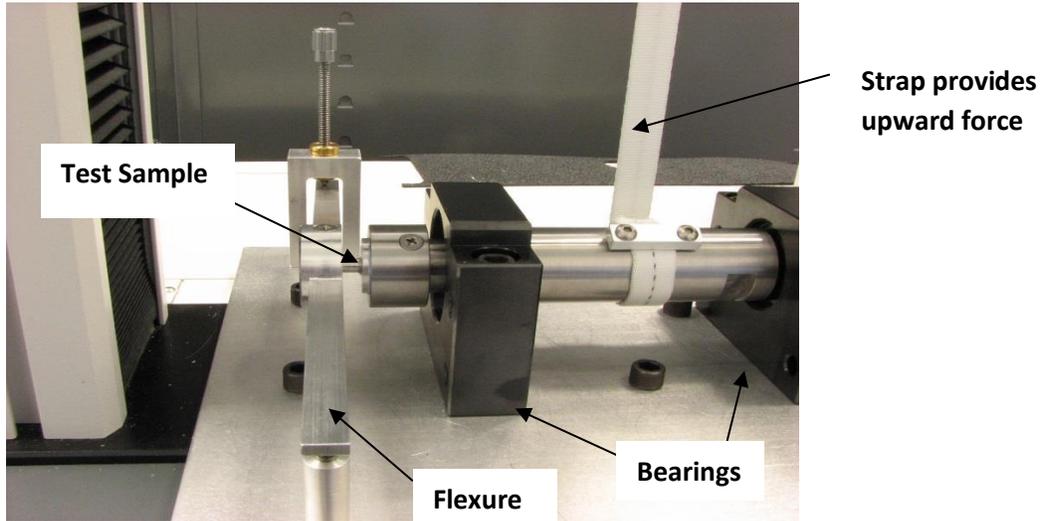

**Figure 55.  Shear test fixture.**

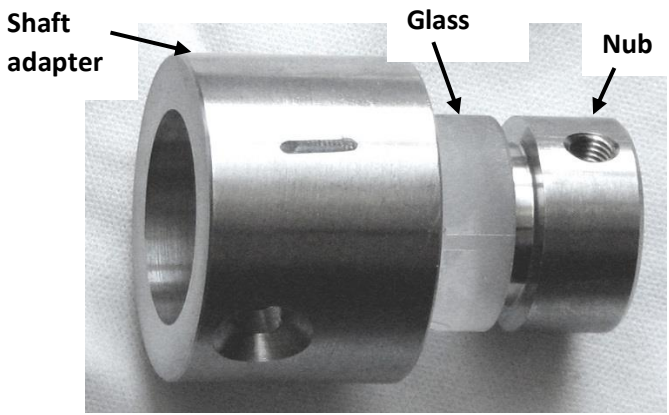

**Figure 56.  Test sample consisting of glass bonded to a metal nub.**

**Table 8. Shear test results**

| Sample | Glass Material | Nub Material | Diameter (mm) | 20 °C Shear (MPa) | -13 °C Shear (MPa) | 20 °C Shear Oil Soaked (MPa) |
|---|---|---|---|---|---|---|
| 1 | BAL15Y | Ti6A14V | 19.0 | 11.7* | 11.7* | |
| 2 | FSL5Y | Ti6A14V | 17.3 | 15.7* | 15.7* | |
| 3 | PBM2Y | Ti6A14V | 17.8 | 14.4* | 14.4* | |
| 4 | PBL6Y | Ti6A14V | 16.8 | 17.2* | 17.2* | |
| 5 | BAL35Y | Kovar | 10.2 | 43.9 | | 48.0 |
| 6 | BSM51Y | Kovar | 12.7 | 39.6* | 39.6* | |
| 7 | CAF2 | 304L SS | 14.2 | 23.3 | 25.7 | 28.2* |
| 8 | S-FPL51Y | 4150 Steel | 11.7 | 42.5 | 44.6 | 50.1 |
| 9 | NaCl | NaCl | 12.7 | 15.2 | 12.5 | |

*Did not break under 890 N vertical load

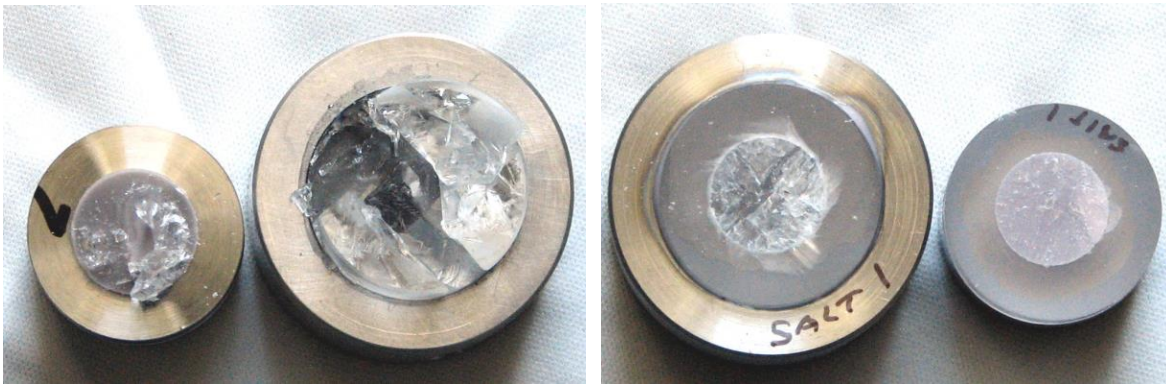

**Figure 57. (Left) Failure of sample 7: CaF2 bonded to a 304L SS nub. (Right) Failure of sample 9: NaCl bonded to NaCl.**

**Appendix 2 – Laser Slit Mask Cutter**

The slit masks for Binospec are cut from 0.25 mm thick 304 stainless steel sheets painted with Aeroglaze Z306 flat black on the side facing the spectrograph to suppress reflections. Five mask pairs can be cut from an 0.61 m square sheet. We selected a StencilLaser G6080 from LPKF Inc. to cut the masks. The G6080 uses a fiber laser operating at ~1 µm wavelength and can cut a typical sheet of masks in about 30 minutes. Figure 58 shows a typical 234 mm x 105 mm slit mask. The cut quality is remarkably good (Figure 59).

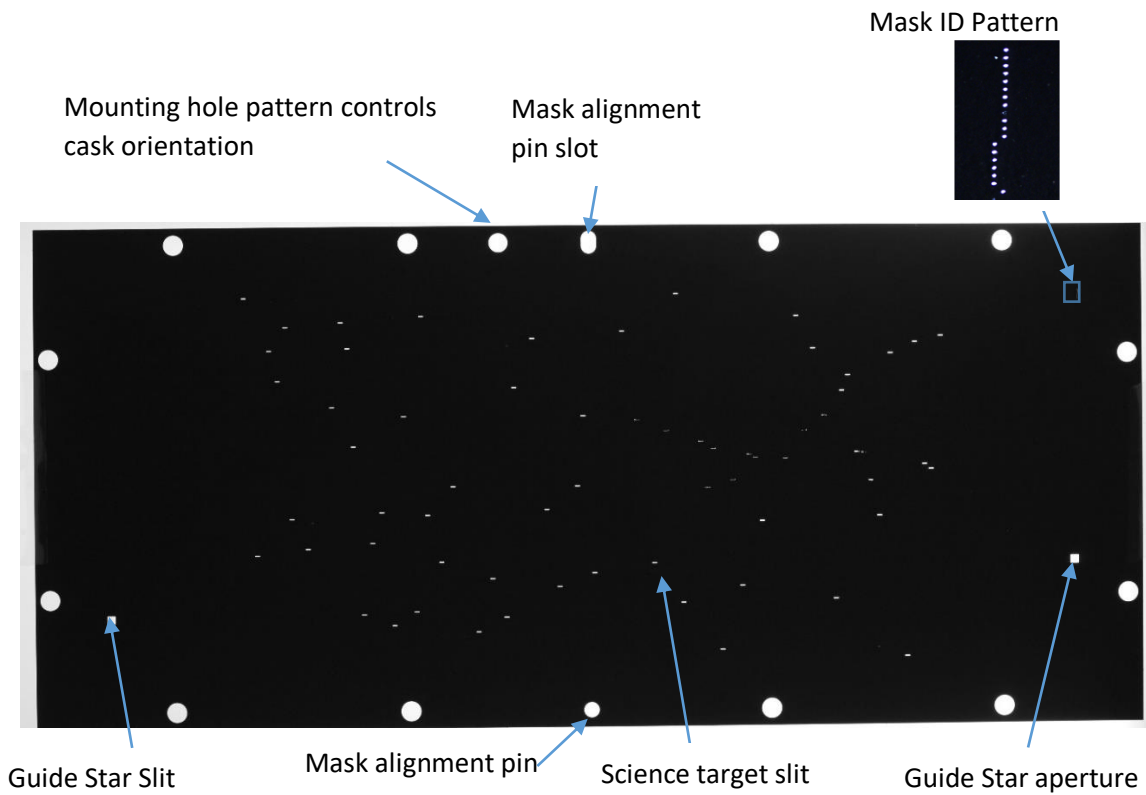

**Figure 58. Typical slit mask used in Binospec**

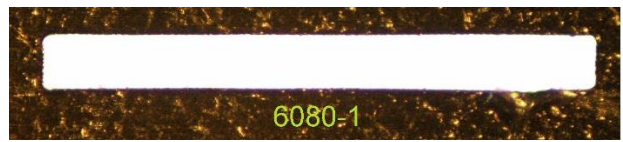

**Figure 59. Closeup of an 85µm x 1000µm individual slit.**

Quality control on the slit masks is performed by placing the slit mask on an overhead projector and viewing the image on a screen. Alternatively, we use a high resolution video camera to display a magnified image on a large monitor.

The slit mask cutting machine will be located just below the summit of the MMT at the Common Building to allow rapid response mask cutting. The ground floor of the building is being remodelled to provide a climate controlled room with a temperature of 21°C ± 2°C and humidity between 30% and 60%.